\documentclass[12pt]{article} 

\usepackage[utf8]{inputenc}
\usepackage[T1]{fontenc} 
\usepackage[english]{babel}           

\usepackage[pdftex]{graphicx}
\usepackage{epsfig}
\usepackage{graphicx}
\usepackage{comment}
\usepackage{latexsym}
\usepackage{hyperref}
\usepackage{amsmath}
\usepackage{mathrsfs}
\usepackage[usenames, dvipsnames]{color}
\usepackage{amsbsy}
\usepackage{amssymb}
\usepackage{amsthm}
\usepackage{amsfonts}
\usepackage{cite}
\usepackage{enumitem}
\usepackage{xcolor}
\usepackage{color}
\usepackage{mathtools}
\usepackage{caption} 
\usepackage{subcaption} 

\usepackage{diagbox}

\newcommand{\be}[0]{\begin{equation}}
\newcommand{\ee}[0]{\end{equation}}

\renewcommand{\thefootnote}{\fnsymbol{footnote}}

\newcommand{\R}{\mathbb{R}}
\newcommand{\Z}{\mathbb{Z}}
\renewcommand{\natural}{\mathbb{N}}
\newcommand{\complex}{\mathbb{C}}
\newcommand{\Ka}{K{\"a}hler }
\renewcommand{\O}{{\cal O}}
\renewcommand{\Re}{{\rm Re}\,}
\renewcommand{\Im}{{\rm Im}\,}

\newcommand{\cc}{\textrm{c.c.}}

\newcommand{\where}{\mbox{where}}

\renewcommand{\and}{\mbox{and}}

\newcommand{\esp}{\phantom{\!\!\overset{\displaystyle |}{.}}}
\newcommand{\espD}{\phantom{\!\!\underset{\displaystyle |}{\cdot}}}

\newcommand{\bm}{\boldmath} 

\newcommand{\N}{{\cal N}}
\newcommand{\A}{{\cal A}}
\newcommand{\B}{{\cal B}}
\renewcommand{\S}{{\cal S}}
\newcommand{\T}{{\cal T}}
\newcommand{\U}{{\cal U}}
\newcommand{\C}{{\cal C}}
\newcommand{\cR}{{\cal R}}

\newcommand{\bi}{{\bar \imath}}
\newcommand{\bj}{{\bar \jmath}}

\catcode`\@=11
\def\marginnote#1{}
\newcount\hour
\newcount\minute
\newtoks\amorpm
\hour=\time\divide\hour by60 \minute=\time{\multiply\hour by60
\global\advance\minute by-\hour}
\edef\standardtime{{\ifnum\hour<12 \global\amorpm={am}%
        \else\global\amorpm={pm}\advance\hour by-12 \fi
        \ifnum\hour=0 \hour=12 \fi
        \number\hour:\ifnum\minute<10 0\fi\number\minute\the\amorpm}}
\edef\militarytime{\number\hour:\ifnum\minute<10 0\fi\number\minute}
\def\draftlabel#1{{\@bsphack\if@filesw {\let\thepage\relax
   \xdef\@gtempa{\write\@auxout{\string
      \newlabel{#1}{{\@currentlabel}{\thepage}}}}}\@gtempa
   \if@nobreak \ifvmode\nobreak\fi\fi\fi\@esphack}
        \gdef\@eqnlabel{#1}}
\def\@eqnlabel{}
\def\@vacuum{}
\def\draftmarginnote#1{\marginpar{\raggedright\scriptsize\tt#1}}
\def\draft{\oddsidemargin -.2truein
        \def\@oddfoot{\sl preliminary draft \hfil
        \rm\thepage\hfil\sl\today\quad\militarytime}
        \let\@evenfoot\@oddfoot \overfullrule 3pt
        \let\label=\draftlabel
        \let\marginnote=\draftmarginnote
   \def\@eqnnum{(\theequation)\rlap{\kern\marginparsep\tt\@eqnlabel}%
\global\let\@eqnlabel\@vacuum}  }
\def\thebibliography#1{
\vskip 0.5cm \centerline{\bf \Large References}
\list{
[\arabic{enumi}]}{\settowidth\labelwidth{[#1]}
\leftmargin\labelwidth
\advance\leftmargin\labelsep
\usecounter{enumi}}
\def\newblock{\hskip .11em plus .33em minus .07em}
\sloppy\clubpenalty4000\widowpenalty4000
\sfcode`\.=1000\relax}

\renewcommand{\theequation}{\arabic{section}.\arabic{equation}}
\renewcommand{\section}{\setcounter{equation}{0}\@startsection
{section}{1}{0mm}{-\baselineskip}{0.5\baselineskip} {\normalfont\Large\bfseries}}
\renewcommand{\subsection}{\@startsection
{subsection}{2}{0mm}{-\baselineskip}{0.5\baselineskip} {\normalfont\large\bfseries}}
\renewcommand{\subsubsection}{\@startsection
{subsubsection}{3}{0mm}{-\baselineskip}{0.5\baselineskip}
{\normalfont\normalsize\slshape}}

\begin{document}


\begin{titlepage}
\begin{flushright}
CPHT-RR015.032019, March 2019
\vspace{1.5cm}
\end{flushright}
\begin{centering}
{\bm\bf \Large HAGEDORN-LIKE TRANSITION AT HIGH \\ 
\vspace{.2cm} SUPERSYMMETRY BREAKING SCALE}

\vspace{7mm}

 {\bf Herv\'e Partouche and Balthazar de Vaulchier}

 \vspace{4mm}

{Centre de Physique Th\'eorique, Ecole Polytechnique,\footnote{Unit\'e  mixte du CNRS et de l'Ecole Polytechnique, UMR 7644.} \\ F--91128 Palaiseau, France\\ \textit{herve.partouche@polytechnique.edu, balthazar.devaulchier@ens.fr}}

\end{centering}
\vspace{0.1cm}
$~$\\
\centerline{\bf\Large Abstract}\\
\vspace{-1cm}

\begin{quote}

\hspace{.6cm} 

We consider phase transitions occurring in four-dimensional heterotic orbifold models, when the scale of spontaneous breaking of $\N=1$ supersymmetry is of the order of the string scale. The super-Higgs mechanism is implemented by imposing distinct boundary conditions for bosons and fermions along an internal circle of radius $R$. Depending on the orbifold action, the usual scalars becoming tachyonic when $R$ falls below the Hagedorn radius may or may not be projected out of the spectrum. In all cases, infinitely many other scalars, which are pure Kaluza-Klein or pure winding states along other internal directions, become tachyonic in subregions in moduli space. We derive the off-shell tree level effective potential that takes into account these potentially tachyonic modes. We show that when a combination of the usual tachyons survives the orbifold action, it is the only degree of freedom that actually condenses. 

\end{quote}

\end{titlepage}
\newpage
\setcounter{footnote}{0}
\renewcommand{\thefootnote}{\arabic{footnote}}
 \setlength{\baselineskip}{.7cm} \setlength{\parskip}{.2cm}

\setcounter{section}{0}


\section{Introduction}

At finite temperature $T$, in a system with an exponential growth of degrees of freedom as a function of mass, the canonical  partition function develops a divergence above the so-called Hagedorn temperature $T_{\rm H}$~\cite{Hage1,Hage2,Hage4}.  
In the context of closed string theory, because modular invariance exchanges ultraviolet and infrared limits, the asymptotic behavior of the density of states and  the light mass spectrum are connected~\cite{Axenides-et-al,Kuta-Seib}.\footnote{This is also true in type I string theory, due to the presence of the closed string sector.} In particular, implementing finite temperature by compactifying Euclidean time on a circle $S^1(R_0)$ of circumference $2\pi R_0=1/T$ with different boundary conditions for bosons and fermions, one observes that 2 real scalars, with winding numbers along $S^1(R_0)$ and generically massive, become tachyonic when the radius $R_0$ falls below $R_{\rm H}=1/(2\pi T_{\rm H})$~\cite{Hage5,Hage6}. This is an indication that the breakdown of the canonical formalism does not result from any pathology of the system, but rather signals an instability occurring at $T=T_{\rm H}$. In fact, tachyons play the role of order parameters whose condensations bring the system in new, stable vacua, as in the Higgs mechanism~\cite{Atick-Witten,AK,ADK,Sen-tach}.\footnote{Finite temperature can also be introduced in a ``non-democratic'' way, namely for all states whose masses are below the string scale~\cite{gravito-magnetic}. In that case, a maximal temperature of order of the string scale exists, there are no tachyonic instabilities, and interesting scenarios can emerge~\cite{cosmo1, cosmo2,cosmo3,cosmo4}.}

At zero temperature, other phase transitions occur in string theory, when a total spontaneous breaking of supersymmetry is induced by a stringy version~\cite{SSstring1,SSstring2,SSstring3,SSstring4,KR,open1_ss,open2_ss,open3_ss,open4_ss,open5_ss} of the Scherk-Schwarz mechanism~\cite{SS1,SS2,SS3}.  In the simplest case, space-time bosons and fermions obey different boundary conditions along an internal (rather than temporal) circle $S^1(R)$ of radius $R$, which induces a supersymmetry breaking at a scale $m_{3\over 2}=1/(2R)$. In toroidal compactifications, instabilities technically similar to the Hagedorn case occur when $R$ reaches $R_{\rm H}$. Bellow this value, 2 real scalars with non-trivial winding numbers along $S^1(R)$ are tachyonic and condense. However, new phenomena may be encountered in orbifold models. This is the case when the above scalars are projected out by the modding action~\cite{SNSM2}. One may think that because tachyons are not allowed in twisted sectors, such models may be tachyon free everywhere in moduli space. However, this conclusion turns out to be incorrect for the following reason. Without orbifold action, the scalars tachyonic for $R<R_{\rm H}$ admit infinite towers of pure Kaluza-Klein (KK) states propagating along internal directions other than $S^1(R)$. Similarly, they  admit infinite towers of pure winding modes wrapped along internal directions other than $S^1(R)$. For large (low) enough volume of these directions, a finite number of KK (winding) modes are therefore tachyonic. In fact, even when the volume of the extra directions is of order~1 in string units,\footnote{Throughout this paper we take $\alpha'=1$.} it turns out that non-trivial KK or winding modes can  be tachyonic. In the descendent orbifold model, because invariant combinations of such potentially tachyonic states survive, there is always a phase in moduli space where a condensation takes place. 

In the present work, we focus on the simplest case, where a single combination of the usual tachyonic states considered in the literature -- i.e. with non-trivial quantum numbers along $S^1(R)$ only -- survives the orbifold action. We take into account other scalars, with identical charges along $S^1(R)$ but non-trivial momenta or winding numbers along another internal direction. These modes can  be tachyonic in more restricted regions in moduli space. The question we ask is whether there exists a multiphase diagram associated with various patterns of condensations, and associated with different stable vacua. We find that all of the condensation is actually supported by the tachyon that has trivial quantum numbers along the directions transverse to $S^1(R)$. In other words, there is a unique Hagedorn-like phase, which is delimited by the usual boundary $R=R_{\rm H}$. Note that this assumes that the Scherk-Schwarz direction is a factorized circle in the internal space. When the  internal  metric and antisymmetric tensor are generic, the boundary of the Hagedorn-like phases  are much more involved. Moreover, when the orbifold action forces all potentially tachyonic states to have non-trivial quantum numbers in the directions transverse to the Scherk-Schwarz circle, the boundaries of the Hagedorn-like phases as well as the properties of the associated vacua are drastically different. However, these generalizations will be analyzed in subsequent work. 

To figure out phase transitions between string models defined in first quantized formalism, the suitable framework should be string field theory~\cite{SFT1,SFT2,SFT3,SFT4}. However, such an analysis being equivalent to describing the vacuum structure of the theory,  the problem may be tackled within an effective field theory, valid at low energy. Such a description can be determined from our knowledge of the phase associated with the initial orbifold compactification. The latter describes a 
super-Higgs mechanism in Minkowski space, with an arbitrary scale $m_{3\over 2}$ of supersymmetry breaking. Hence, it  is a no-scale supergravity~\cite{noscale}, which takes into account all light and potentially tachyonic degrees of freedom. The key point is that the supergravity action is valid {\rm off-shell}. Therefore, it captures other vacua characterized by non-trivial condensates developed in regions in moduli space where tachyonic instabilities take place. Notice that our use of the word ``vacuum'' is cavalier in the sense that the tachyon  condensation lowers the potential of the theory to negative values, which yields a dilaton tadpole. As a result, the new supergravity phase may describe a non-critical string at low energy, with linear dilaton background~\cite{AK,ADK}. 

In Sect.~\ref{Stach}, we consider as a starting point the heterotic string compactified on  $T^2\times T^2\times T^2$. A Scherk-Schwarz mechanism responsible for the $\N=4\to \N=0$ spontaneous  breaking of supersymmetry in 4 dimensions is implemented along one direction, $X^4$, of the first internal $T^2\equiv S^1(R_4)\times S^1(R_5)$~\cite{SSstring1,SSstring2,SSstring3,SSstring4,KR}. We determine the regions of the plan $(R_4,R_5)$, where scalars with non-trivial momentum and/or winding numbers along $X^4$ and $X^5$ are tachyonic. We then introduce a $\Z_2\times \Z_2$ orbifold action and analyze the conditions for a tachyonic mode with trivial quantum numbers along $S^1(R_5)$ to survive. We stress that the latter is accompanied by an infinite number of potentially tachyonic KK (or winding) modes propagating along (or wrapped around) $S^1(R_5)$. 
In Sect.~\ref{cond}, we derive the tree level effective potential that depends on all of these scalars. This may be done in the framework of $\N=1$ supergravity~\cite{N=1-1,N=1-2}. However, because all degrees of freedom of interest arise in the untwisted sector of the $\Z_2\times \Z_2$ orbifold action, we find convenient to derive the potential by applying a suitable truncation of $\N=4$ gauged supergravity~\cite{sugra1,sugra2,sugra2bis,sugra3,sugra4,sugraN=4}. In this formalism, the gauging is determined by imposing the mass spectrum in the no-scale supergravity phase to reproduce its string counterpart. The {\em off-shell} tree level bosonic action is found to be invariant under the modified T-duality $R_4\rightarrow 1/(2R_4)$. This is consistent with the fact that this transformation (accompanied with a change of chirality for the fermions) is a symmetry of the 1-loop partition function of the initial string model, and thus a symmetry of the {\em on-shell} 1-loop effective potential,  in the no-scale phase.
It is  straightforward to minimize the tree level potential to find that in the case under study, the only mode that condenses in the Hagedorn-like phase  is the tachyon that has quantum numbers along the Scherk-Schwarz direction $S^1(R_4)$ only. Because the  new vacuum lies at the self-dual radius $\langle R_4 \rangle = 1/\sqrt{2}$, the T-duality $R_4\rightarrow 1/(2R_4)$ is not spontaneously broken. Finally, our conclusions and perspectives can be found in Sect.~\ref{cl}. 


\section{Tachyonic phases}
\label{Stach}

In this section, our aim is to characterize regions in moduli space where one or several generically massive states become tachyonic for sufficiently large supersymmetry breaking scale. The resulting condensation phenomenon will be discussed in Sect.~\ref{cond}.


\subsection{Towers of KK or winding tachyonic states}

\begin{figure}[!h]
\centering
\includegraphics[scale=1]{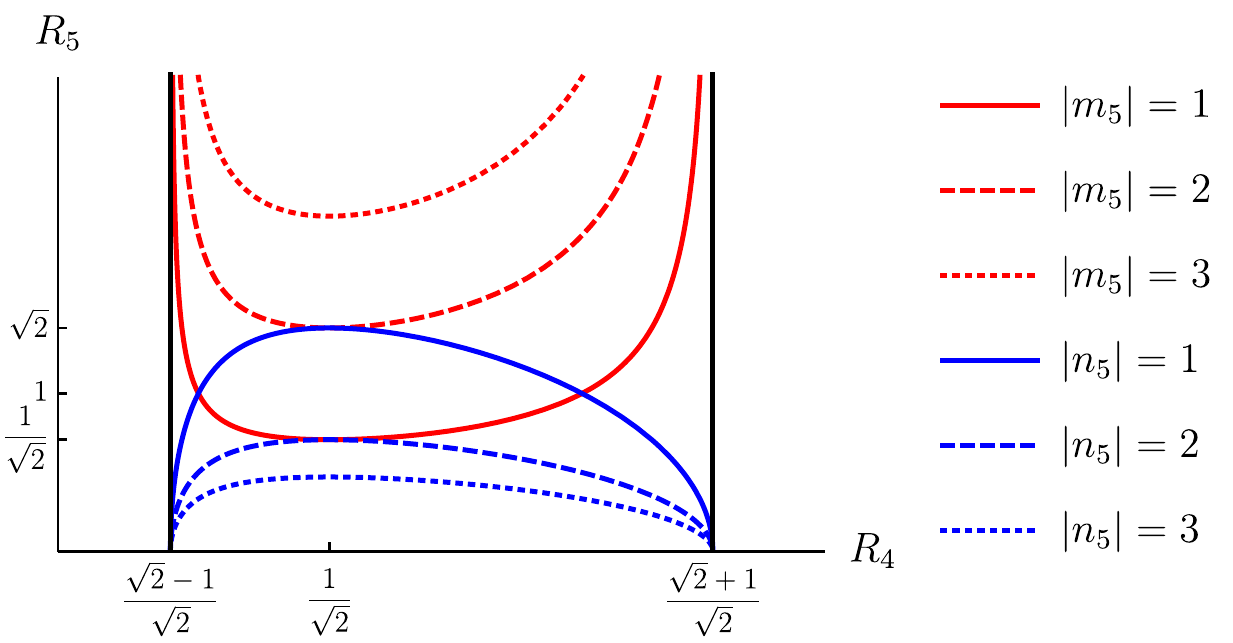}
\caption{\footnotesize {\em Boundary curves of the regions of the plan $(R_4,R_5)$ where KK or winding modes along $S^1(R_5)$ are tachyonic in the parent model, which realizes the $\N=4\to \N=0$ spontaneous breaking. The remaining quantum numbers of these states are $2m_4=-n_4=\epsilon$ and $m_6=n_6=\dots=m_9=n_9=0$.}}
\label{domains}
\end{figure}

In the present work, we consider  the heterotic string  compactified on the orbifold  $T^6/(\Z_2\times \Z_2)$. For simplicity, the analysis is restricted to the case where the internal $T^6$ is of the form~
\be
S^1(R_4)\times S^1(R_5)\times T^2\times T^2,
\label{back}
\ee
i.e. with first 2-torus factorized into two circles of radii $R_4$ and $R_5$. The spontaneous breaking of  $\N=1$ supersymmetry is implemented along the compact direction $X^4$, by a stringy version~\cite{SSstring4, KR} of the Scherk-Schwarz mechanism~\cite{SS1,SS2}. The zero point energy in the twisted sectors being non-negative, tachyons can only arise in the untwisted sector. Before $\Z_2\times \Z_2$ projection, the associated 1-loop partition function takes the following form,
\begin{align}
&Z =  {R_4\over \sqrt{\Im \tau}} \sum_{n_4,\tilde m_4}e^{-{\pi R^2_4\over \Im \tau}|\tilde m_4+n_4\tau|^2}\, {\Gamma_{(1,1)}(R_5) \Gamma_{(2,2)} \Gamma_{(2,2)} \Gamma_{(0,16)}\over \tau_2^2\, \eta^8\bar \eta^{24}}\, {1\over 2}\sum_{\alpha,\beta} (-1)^{\alpha+\beta+\alpha\beta}\, {\theta[^\alpha_\beta]^4\over \eta^4}\,\C\big[{}^{\alpha;\, n_4}_{\beta\, ;\tilde m_4}\big] ,\nonumber \\
&\where \quad \C\big[{}^{\alpha;\, n_4}_{\beta\, ;\tilde m_4}\big] =  (-1)^{\tilde m_4 \alpha+n_4\beta+\tilde m_4n_4}.\esp
\label{mod}
\end{align}
In our notations, $\tau$ is the Teichm\"uller parameter of the genus-1 Riemann surface and our definitions of the Jacobi modular forms $\theta[^\alpha_\beta]$ and Dedekind  function $\eta$ are as follows,
\be
\theta[^\alpha_\beta](\tau)=\sum_{N} q^{{1\over 2}(N-{\alpha\over 2})^2}\, e^{-i\pi \beta(N-{\alpha\over 2})},\quad \eta(\tau)=q^{1\over 24}\prod_{k\ge 1}(1-q^k),\quad q=e^{2i\pi \tau},
\label{def}
\ee
where $N\in\Z$.
The lattices of bosonic zero-modes associated with $S^1(R_5)$, the $T^2$'s and the extra right-moving coordinates are denoted $\Gamma_{(p,q)}$, while that  associated to $S^1(R_4)$ is written in Lagrangian form, where $n_4, \tilde m_4\in\Z$. The conformal blocks arising from the left-moving worldsheet fermions depend on the spin structures $\alpha,\beta\in\{0,1\}$. The latter are coupled to the $S^1(R_4)$ lattice of zero modes by the ``cocycle'' $\C\big[{}^{\alpha;\, n_4}_{\beta\, ;\tilde m_4}\big]$~\cite{KR}. To see that this sign breaks spontaneously supersymmetry, one can switch to a Hamiltonian formulation obtained by Poisson summation over $\tilde m_4$. One obtains
\be
\begin{aligned}
Z =    {\Gamma_{(2,2)}\, \Gamma_{(2,2)}\, \Gamma_{(0,16)}\over \tau_2^2\, \eta^8\bar \eta^{24}} \sum_{n_5,m_5} \Gamma_{m_5,n_5}&(R_5)\bigg\{\sum_{n_4\text{even}, m_4} \left(\Gamma_{m_4,n_4}(R_4) V_8-\Gamma_{m_4+{1\over 2},n_4}(R_4)S_8\right)\\
&\;\;\;\;- \sum_{n_4\text{odd}, m_4} \left(\Gamma_{m_4+{1\over 2},n_4} (R_4)O_8-\Gamma_{m_4,n_4}(R_4)C_8\right)\bigg\},
\end{aligned}
\ee
where we denote for $I\in\{4,5\}$ and $\kappa=0,1$ 
\be
\Gamma_{m_I+{\kappa\over 2},n_I}(R_I) = q^{{1\over 2}p_{I\rm L}^2}\, \bar q^{{1\over 2}p_{I\rm R}^2},\quad p_{I\underset{\scriptstyle \rm R}{\rm L}}={1\over \sqrt{2}}\Big({m_I+{\kappa\over 2}\over R_I}\pm n_IR_I\Big), \quad m_I,n_I\in\Z, 
\ee
while $SO(8)$ affine characters are defined as 
\be
O_{8}={\theta[{}^0_0]^4+\theta[{}^0_1]^4\over 2\eta^4} , \quad V_{8}={\theta[{}^0_0]^4-\theta[{}^0_1]^4\over 2\eta^4} , \quad 
S_{8}={\theta[{}^1_0]^4+\theta[{}^1_1]^4\over 2\eta^4},
\quad C_{8}={\theta[{}^1_0]^4-\theta[{}^1_1]^4\over 2\eta^4}.
\ee
Comparing the lattice dressing of the characters $V_8$ and $S_8$, the supersymmetry breaking scale (or gravitino mass) in $\sigma$-model frame is found to be
\be
m_{3\over 2}={1\over 2R_4}.
\label{m32}
\ee
If the sign $(-1)^{n_4\beta}$ present in the cocycle reverses the GSO projection in the odd $n_4$ winding sector, the associated characters $O_8,C_8$ yield states heavier than the string scale when  $\mbox{$R_4\gg 1$}$. Due to the  T-duality  
\be
(R_4, S_8,C_8)\longrightarrow \Big({1\over 2 R_4}, C_8,S_8\Big)
\ee
satisfied by $Z$, the characters $O_8$ and $S_8$ also lead to very heavy modes when $R_4\ll 1$. However, the leading term of the $q,\bar q$-expansion 
\be
{O_8\over \eta^8\bar \eta^{24}}={1\over q^{1\over 2}\bar q}(1+\O(q)+\O(\bar q))
\ee
can yield tachyonic scalars, when $R_4$ is of order 1. 
Denoting $m_I,n_I$ the momentum and winding numbers along the internal directions $X^I$, $I\in\{5,\dots,9\}$, the level matching condition at this oscillator level reads
\be
{1\over 2}+\Big(m_4+{1\over 2}\Big)n_4+\sum_{I=5}^9m_In_I=0.
\ee
The physical states that can be tachyonic in the parent model realizing the $\N=4\to \N=0$ spontaneous breaking of supersymmetry turn out to have quantum numbers  
\be
2m_4+1=-n_4=\epsilon, \quad m_5 n_5=\dots=m_9 n_9=0,
\ee
where $\epsilon = \pm 1$. They have non-trivial momentum and winding numbers along the Scherk-Schwarz direction, but are pure KK or winding modes along the remaining internal directions. For instance, the squared masses in $\sigma$-model frame of  those having  $m_6=n_6=\dots=m_9=n_9=0$ are
\be
M^2_{(\epsilon,m_5,0)}={1\over 4R_4^2}+R_4^2-3+\Big({m_5\over R_5} \Big)^2, \qquad M^2_{(\epsilon,0,n_5)}={1\over 4R_4^2}+R_4^2-3+(n_5 R_5 )^2.
\label{massesT}
\ee
The largest tachyonic domain in the plane $(R_4,R_5)$ is obtained for $m_5=n_5=0$,
\be
{\sqrt{2}-1\over \sqrt{2}}={1\over 2R_{\rm H}}<R_4<R_{\rm H}={\sqrt{2}+1\over \sqrt{2}}, \quad\;\;R_5\mbox{ arbitrary},
\label{range}
\ee    
where $R_{\rm H}$ is the Hagedorn radius encountered in heterotic string at finite temperature. However, subregions where additional states are tachyonic also exist, since 
\be
\begin{aligned}
M_{(\epsilon,m_5,0)}^2<0 \quad &\Longleftrightarrow \quad R_{\rm H-}\big|_{m_5\over R_5}<R_4<R_{\rm H+}\big|_{m_5\over R_5}, \quad\;\;\;\; \;\,{|m_5|\over R_5}<\sqrt{2},\espD\\
M_{(\epsilon,0,n_5)}^2<0\quad& \Longleftrightarrow \quad R_{\rm H-}\big|_{n_5 R_5}\;\;<R_4<R_{\rm H+}\big|_{n_5 R_5}, \quad {|n_5|R_5}<\sqrt{2},\espD\\
 \where \quad &R_{\rm H\pm}|_x={1\over 2}\Big[6-2x^2\pm \sqrt{[6-2x^2]^2-4}\Big]^{1\over 2}.
\end{aligned}
\ee
Fig.~\ref{domains} shows in blue (red) the boundaries of the domains $M^2_{(\epsilon,0,n_5)}<0$ ($M^2_{(\epsilon,m_5,0)}<0$), for $|n_5|=1,2,3$ ($|m_5|=1,2,3$).
More and more  modes winding along $S^1(R_5)$ become tachyonic as $R_5$ decreases, while more and more KK modes propagating along $S^1(R_5)$ become tachyonic as $R_5$ increases. By analogy with the finite temperature case~\cite{Hage5,Hage6,Atick-Witten,AK,ADK}, a Hagedorn-like phase transition is expected to occur when $R_4$ enters the range~(\ref{range}). An instability of the initial no-scale model  vacuum~\cite{noscale}  should be signalled  by the condensation of, at least, the tachyonic modes $2m_4=-n_4=\epsilon, m_5=n_5=\dots=m_9=n_9=0$. As seen on Fig.~\ref{domains}, a multi-phase diagram may however exist, with different vacua characterized by various condensed modes.  


\subsection{Why supergravity}

As soon as $R_4$ enters the range~(\ref{range}), implying $M_{(\epsilon,0,0)}^2$ to be negative in the parent model that realises the $\N=4\to \N=0$ breaking, the 1-loop effective potential, which is nothing but the partition function~(\ref{mod}) integrated over the fundamental domain of $SL(2,\Z)$, diverges. The fact that the  quantum potential is ill-defined does not signal some fundamental inconsistency of the theory. Indeed, this means that perturbative quantum corrections should be computed around a new, true vacuum~\cite{Atick-Witten, AK, ADK,Sen-tach}. In the latter, the derivation of the mass spectrum using the  initial string background is not legitimate anymore. Hence, once the modes $2m_4+1=-n_4=\epsilon, m_5=n_5=\dots=m_9=n_9=0$ have already condensed, we should consider as a possibility, rather than a prediction, the condensation of other KK or winding modes along, for instance,  $S^1(R_5)$. This is the reason why we will derive in Sect.~\ref{cond} the {\em off-shell} low energy effective potential associated with all of these potentially condensing degrees of freedom, in order to figure out which of them actually develop non-trivial expectation values.  

In presence of the cocycle $\C$~\cite{KR} in the partition function~(\ref{mod}), the GSO projection being reversed in the odd  $n_4$ winding sector, the (common) statement that the non-supersymmetric model arises as a spontaneous breaking of a supersymmetric theory (i.e. with no cocycle) is not obvious. To see this is the case, let us recall the initial formulation of the stringy Scherk-Schwarz mechanism as a ``coordinate-dependent compactification''~\cite{SSstring2,SSstring4}. In our case of interest, this  amounts to  coupling the lattice $\Gamma_{(1,1)}(R_4)$   to the  boundary conditions of 2 (among 8) real  left-moving worldsheet  fermions. Before implementation of the orbifold action, the choice of these fermions is arbitrary since they all have identical boundary conditions. This leads
\be
{R_4\over \sqrt{\Im \tau}} \sum_{n_4,\tilde m_4} e^{-{\pi R^2_4\over \Im \tau}|\tilde m_4+n_4\tau|^2} \, e^{-i\pi n_4 e (\tilde m_4 e-\beta)}\, \theta\big[{}^{\alpha-2n_4e}_{\beta-2\tilde m_4 e}\big],
\label{defo}
\ee
where $e$ is the coupling defining a deformation of the supersymmetric model. It can be chosen in the range $-1<e\le 1$, due to the symmetry $e\to e+2$. The precise form of the deformation is motivated by the fact that the modular transformations of the above conformal blocks turn out to be independent of $e$. Using the definition~(\ref{def}), a Poisson summation over $\tilde m_4$ yields 
\be
\sum_{n_4,m_4,N} e^{-i\pi \beta(N-{\alpha\over 2})}\, q^{{1\over 2}[p_{4\rm L}^2+(N-{\alpha\over 2}+n_4 e)^2]}\, \bar q^{{1\over 2}p_{4\rm R}^2},
\label{Nq}
\ee
where the generalized momenta  in this formulation take the following form,
\be
p_{4\underset{\scriptstyle R}{L}}={1\over \sqrt{2}}\Big({m_4-e(N-{\alpha\over 2})-{1\over 2}n_4e^2 \over R_4}\pm n_4R_4\Big).
\label{pq}
\ee 
We see that both the GSO projection (the $\beta$-dependent phase) and the level-matching condition are  independent of $e$, since 
\be
{1\over 2}\!\left[p_{4\rm L}^2+\Big(N-{\alpha\over 2}+n_4 e\Big)^2-p_{4\rm R}^2\right]\!=m_4n_4+{1\over 2}\Big(N-{\alpha\over 2}\Big)^2.
\ee
Therefore,  there is a one-to-one correspondence between the states of the supersymmetric and deformed theories. The mass spectrum, however, depends on $e$. For instance, the masses of the 4 gravitini (or their surviving combination after $\Z_2\times \Z_2$ projection) are $m_{3\over 2}=|e|/(2R_4)$. However, consistency of the heterotic worldsheet theory imposes the supercurrent to be conserved, which forces the deformation to be quantized~\cite{SSstring2,SSstring4,KR}.\footnote{Similar deformations of the boundary conditions of right-moving worldsheet degrees of freedom in fermionic language are not quantized. They are interpreted as non-trivial continuous Wilson lines responsible for a Higgs mechanism~\cite{421,QNSR2}.}   Taking  $e=1$, not only the gravitino mass~(\ref{m32}) is recovered, since the properties of the Jacobi modular forms can be used to rewrite the conformal blocks~(\ref{defo}) in the following form,
\be
{R_4\over \sqrt{\Im \tau}} \sum_{n_4,\tilde m_4} e^{-{\pi R^2_4\over \Im \tau}|\tilde m_4+n_4\tau|^2} \, \theta\big[{}^{\alpha}_{\beta}\big]\, e^{-i\pi e(\tilde m_4 \alpha-n_4\beta+e\tilde m_4n_4)},
\label{co1}
\ee
i.e. with the cocycle introduced in Eq.~(\ref{mod}). Because the consistent quantum field theories of massive spin~${3\over 2}$ particles are supergravities realizing the super-Higgs mechanism, the low energy effective field theory associated with the untwisted sector of the $\Z_2\times \Z_2$ model can be described by an $\N=4$ gauged supergravity~\cite{sugra1,sugra2,sugra2bis,sugra3,sugra4,sugraN=4}, with suitable truncation. 




\subsection{$\Z_2\times \Z_2$ projection}
\label{secZ}

We would like now to implement the orbifold action whose generators are defined as 
\be
\begin{aligned}
\boldsymbol{G_1}: \quad (X^4,X^5,X^6,X^7,X^8,X^9)\longrightarrow (X^4,X^5,-X^6,-X^7,-X^8,-X^9),\\
\boldsymbol{G_2}: \quad(X^4,X^5,X^6,X^7,X^8,X^9)\longrightarrow (-X^4,-X^5,X^6,X^7,-X^8,-X^9).
\end{aligned}
\ee
We are going to see that there are conditions for the coordinate-dependent compactification to be implemented consistently and that the surviving set of potentially tachyonic modes is model-dependent. 

Let us first implement the $\Z_2$ action generated by $\boldsymbol{G_1}$, which breaks $\N=4$ to $\N=2$. Because the boundary conditions of the left-moving worldsheet fermions $\psi^6$, $\psi^7$, $\psi^8$, $\psi^9$ are initially identical, we may deform those of any two of them to implement the coordinate-dependent compactification. For instance, choosing $\psi^6$, $\psi^7$, the relevant conformal blocks are 
\be
{R_4\over \sqrt{\Im \tau}} \sum_{n_4,\tilde m_4} e^{-{\pi R^2_4\over \Im \tau}|\tilde m_4+n_4\tau|^2} \, e^{-i\pi n_4 e (\tilde m_4 e-(\beta+G_1))}\, \theta\big[{}^{\alpha}_{\beta}\big]^2\theta\big[{}^{\alpha+H_1-2n_4e}_{\beta+G_1-2\tilde m_4 e}\big]\theta\big[{}^{\alpha+H_1}_{\beta+G_1}\big](-1)^{\xi(\tilde m_4H_1-n_4G_1)},
\label{defo2}
\ee
where we have written explicitly the dependence on the quantum number $H_1\in\{0,1\}$, which labels the untwisted and twisted sectors of the $\Z_2$ modding, as well as the dependance on $G_1 \in\{0,1\}$, which signals the insertion ${\boldsymbol{G_1}}^{G_1}$ in the supertraces. In the above formula, $\xi=0$ or 1 defines two distinct choices of discrete torsions, which are allowed by modular invariance. For $e=1$, the expression can be rewritten in terms of a modified cocycle 
\be
\begin{aligned}
&{R_4\over \sqrt{\Im \tau}} \sum_{n_4,\tilde m_4} e^{-{\pi R^2_4\over \Im \tau}|\tilde m_4+n_4\tau|^2} \,  \theta\big[{}^{\alpha}_{\beta}\big]^2\theta\big[{}^{\alpha+H_1}_{\beta+G_1}\big]^2\,  \C\big[{}^{\alpha;\, n_4}_{\beta\, ;\tilde m_4}\big]  \, \C'\big[{}^{H_1;\, n_4}_{G_1 ;\tilde m_4}\big]  , \\
&\where \quad \C'\big[{}^{H_1;\, n_4}_{G_1 ;\tilde m_4}\big]   =  (-1)^{(1-\xi)(\tilde m_4H_1-n_4G_1)}.\esp
\end{aligned}
\label{co2}
\ee
With the cocycle prescription $\C$ alone, we have seen that the potentially  tachyonic modes are untwisted, $H_1=0$, and have odd winding number $n_4$. In such a sector, because $\C'$ reduces to $(-1)^{(1-\xi)G_1}$, the presence of $\C'$ in the blocs~(\ref{co2}) modifies the $\Z_2$ projector as follows,
\be
{1\over 2}\sum_{G_1=0}^1 {\boldsymbol{G_1}}^{G_1}=  {1+{\boldsymbol{G_1}}\over 2}\quad \longrightarrow\quad
 {1\over 2}\sum_{G_1=0}^1 (-1)^{(1-\xi)G_1} {\boldsymbol{G_1}}^{G_1}  = {1-(-1)^\xi{\boldsymbol{G_1}}\over 2},
\label{proj}
\ee
which has important consequences. 

To see this explicitly, let us use the notations of Eq.~(\ref{pq}), where the potentially tachyonic modes have quantum numbers  $m_4=-n_4=-Q=\epsilon$, where $Q\equiv -N+{\alpha\over 2}$ are the eigenvalues of the generator of the $SO(2)$ affine algebra associated with the normal-ordered conserved current $:\!\psi^6\psi^7\!:$. In complex notations, the affine generator is defined as
\be
Q={1\over 2\pi}\int_0^{2\pi} d\sigma^1 :\left({\psi^6+i\psi^7\over \sqrt{2}}\right)^\dag{\psi^6+i\psi^7\over \sqrt{2}}:,
\ee
where $\sigma^1$ is the coordinate along the string. 
The remaining non-trivial quantum numbers of the potentially tachyonic modes satisfy $m_In_I=0$, $I\in\{5, \dots, 9\}$. For instance, those with vanishing winding numbers along $S^1(R_5)\times T^2\times T^2$ have equal left- and right-moving momenta $p_{I\rm L}=p_{I\rm R}$, $I\in\{5,\dots, 9\}$. Under the $\Z_2$ action, they transform as\footnote{In the following, $e^{ip_{I\rm L}X^I_{\rm L}+ip_{I\rm R}X^I_{\rm R}}|0\rangle_{\rm NS}\otimes |\tilde 0\rangle$ stands for $|p_{\rm L}\rangle_{\rm NS}\otimes |\widetilde{p_{\rm R}}\rangle$, and $X^I=X^I_{\rm L}+X^I_{\rm R}$.} 
\be
\begin{aligned}
&{\psi^6+i\epsilon \psi^7\over \sqrt{2}}\, e^{i\epsilon X^4_{\rm R}}\,e^{ip_{5\rm L}X^5}\, e^{i\sum_{I=6}^9 p_{I\rm L}X^I} |0\rangle_{\rm NS}\otimes |\tilde 0\rangle\\
\longrightarrow\;\; -(-1)^\xi\, &{\psi^6+i\epsilon \psi^7\over \sqrt{2}}\, e^{i\epsilon X^4_{\rm R}}\,e^{ip_{5\rm L}X^5}\, e^{i\sum_{I=6}^9 (-p_{I\rm L})X^I} |0\rangle_{\rm NS}\otimes |\tilde 0\rangle,
\end{aligned}
\label{ima}
\ee
where $|0\rangle_{\rm NS}$ and $|\tilde 0\rangle$ are the left-moving NS and right-moving vacua. For the sake of simplicity in the notations, we have set $R_4=1/\sqrt{2}$ in the above expressions, which yields $p_{4\rm L}=0$, $p_{4\rm R}=\epsilon$. We see that there are always invariant linear combinations of states surviving the $\Z_2$ projection. However, the modes with trivial momenta and winding numbers along the twisted directions $X^6$, $X^7$, $X^8$, $X^9$ (and whose masses are given in Eq.~(\ref{massesT}))  exist in the orbifold model only if  $\xi=1$. In that case, we will see in the next section that these states condense in the range~(\ref{range}). On the contrary, these modes are  projected out when $\xi=0$, and the properties of the Hagedorn-like phase and its boundary in moduli space must be drastically different. In fact, all potentially tachyonic modes surviving the $\Z_2$ projection when $\xi=0$ are pure KK modes (or pure winding modes) along one or more directions of $T^2\times T^2$ (and possibly along $S^1(R_5)$). 

Let us apply $\boldsymbol{G_2}$ on the potentially tachyonic modes arising in the $\N=4\to \N=0$ model. For instance, those with vanishing winding numbers along  $S^1(R_5)\times T^2\times T^2$ are transformed as
\be
\begin{aligned}
&{\psi^6+i\epsilon \psi^7\over \sqrt{2}}\, e^{i\epsilon X^4_{\rm R}}\,e^{ip_{5\rm L}X^5}\, e^{i\sum_{I=6}^9 p_{I\rm L}X^I} |0\rangle_{\rm NS}\otimes |\tilde 0\rangle\\
\longrightarrow\;\; &{\psi^6+i\epsilon \psi^7\over \sqrt{2}}\, e^{i(-\epsilon) X^4_{\rm R}}\,e^{i(-p_{5\rm L})X^5}\, e^{i\left(p_{6\rm L}X^6+p_{7\rm L}X^7-p_{8\rm L}X^8-p_{9\rm L}X^9\right)} |0\rangle_{\rm NS}\otimes |\tilde 0\rangle.
\end{aligned}
\ee
Notice that the images {\em are not} potentially tachyonic states, but only some of their non-degenerate (and massive) bosonic superpartners, which have $Q=\epsilon$. In other words, $\boldsymbol{G_2}$ {\em is not} a symmetry of the parent $\N=4\to \N=0$ model. In order to construct a consistent $\Z_2\times \Z_2$ orbifold model, it is however possible to implement the coordinate-dependent compactification with the $SO(2)$ affine generator associated with $\psi^6$, $\psi^8$ instead of $\psi^6$, $\psi^7$. Remember that from the point of view of the first generator $\boldsymbol{G_1}$, this is nothing but an equivalent conventional choice, so that Eq.~(\ref{ima}) becomes
\be
\begin{aligned}
&{\psi^6+i\epsilon \psi^8\over \sqrt{2}}\, e^{i\epsilon X^4_{\rm R}}\,e^{ip_{5\rm L}X^5}\, e^{i\sum_{I=6}^9 p_{I\rm L}X^I} |0\rangle_{\rm NS}\otimes |\tilde 0\rangle\\
\longrightarrow\;\; -(-1)^\xi\, &{\psi^6+i\epsilon \psi^8\over \sqrt{2}}\, e^{i\epsilon X^4_{\rm R}}\,e^{ip_{5\rm L}X^5}\, e^{i\sum_{I=6}^9 (-p_{I\rm L})X^I} |0\rangle_{\rm NS}\otimes |\tilde 0\rangle.
\end{aligned}
\label{rrr}
\ee
However, applying  $\boldsymbol{G_2}$ on the same states, we find
\be
\begin{aligned}
&{\psi^6+i\epsilon \psi^8\over \sqrt{2}}\, e^{i\epsilon X^4_{\rm R}}\,e^{ip_{5\rm L}X^5}\, e^{i\sum_{I=6}^9 p_{I\rm L}X^I} |0\rangle_{\rm NS}\otimes |\tilde 0\rangle\\
\longrightarrow\;\; &{\psi^6+i(-\epsilon) \psi^8\over \sqrt{2}}\, e^{i(-\epsilon) X^4_{\rm R}}\,e^{i(-p_{5\rm L})X^5}\, e^{i\left(p_{6\rm L}X^6+p_{7\rm L}X^7-p_{8\rm L}X^8-p_{9\rm L}X^9\right)} |0\rangle_{\rm NS}\otimes |\tilde 0\rangle.
\end{aligned}
\label{rrrr}
\ee
The images are now potentially tachyonic in the parent $\N=4\to \N=0$ model, and  the latter admits a  $\Z_2$ symmetry generated by $\boldsymbol{G_2}$~\cite{SSstring4,KR,SS3}. Hence, a consistent model realizing the $\N=1\to \N=0$ spontaneous breaking of supersymmetry is obtained by modding by $\Z_2\times\Z_2$. Because the affine $SO(2)$ generator involved in the coordinate-dependent compactification rotates the directions 6 and 8,\footnote{For $e=1$, the latter is a $2\pi$-rotation. This action is however non-trivial in the Ramond sector $\alpha=1$ (space-time fermions)~\cite{KR}.} moduli of the tori $T^2\times T^2$ associated with the directions $6, 7$ and $8, 9$ are lifted. This has been analyzed in supergravity in Ref.~\cite{SS3}, while in the string theory context of Refs~\cite{SSstring4,KR}, the moduli of $T^2\times T^2$ take specific values, at fermionic points. 

From now on, we consider the above model with $\xi=1$.\footnote{In Refs~\cite{SSstring4,KR}, the choice of discrete torsion is $\xi=0$.} The potentially tachyonic states are the linear combinations of modes invariant under  the mappings~(\ref{rrr}), (\ref{rrrr}), as well as their counterparts with winding numbers rather than KK momenta along some of the internal directions $X^5,\dots, X^9$.


\section{Effective gauged supergravity}
\label{cond}

Having identified the states potentially tachyonic, our goal is to derive their off-shell tree level potential and its minima, in order to figure out all different phases. 

Gauged $\N=4$ supergravity in four dimensions  is a theory that couples a gravity multiplet to an arbitrary number $k$ of vector multiplets~\cite{sugra1,sugra2,sugra2bis,sugra3,sugra4,sugraN=4}. The scalar content is a complex scalar $\S$ (related to the string theory axion $\chi$ and dilaton $\phi_{\rm dil}$) and $6k$ real fields that realize a non-linear $\sigma$-model with  target space 
\be
{SU(1,1)\over U(1)}\times {SO(6,k)\over SO(6)\times SO(k)}.
\label{6k}
\ee
Properties of such manifolds are briefly reviewed in the appendix. To describe the potential of the theory, it is however convenient to consider the group quotient $SO(6,k)/SO(k)$~\cite{sugra1,sugra2}. The latter can be parameterized by real scalars $Z^S_a$, $S=1,\dots,6+k$, $a=1,\dots, 6$, satisfying\footnote{In the appendix, $Z^S_a$ is denoted ${M^S}_a$.} 
\be
\forall a,b\in\{1,\dots,6\},\quad \eta_{ST} Z^S_aZ^T_b=-\delta_{ab},
\label{var}
\ee
where $\eta=\mbox{diag}(-1,\dots,-1,1,\dots,1)$, with 6 entries $-1$.
Hence, the $Z^S_a$'s describe all physical field configurations, with an $SO(6)$ redundancy. Given the supermultiplet content, the model is further characterized by the gauging, which implements the non-Abelian nature of the vector bosons arising from the vector multiplets and/or the 6 graviphotons. The gauging amounts to switching on structure constants $f_{RST}$ that are totally antisymmetric in their indices. By supersymmetry, the following  potential in Einstein frame  is generated~\cite{sugra2,sugra2bis}, 
\be
V = {|\Phi|^2\over 4}\,  Z^{RU}Z^{SV}\Big( \eta^{TW} + \frac{2}{3}Z^{TW} \Big) f_{RST}f_{UVW},
\label{V4}
\ee  
where the $Z^{RU}$'s are $SO(6)$-invariant combinations,
\be
Z^{RU}=Z^R_aZ^U_a,
\ee
and the overall factor is related to $\S$ as follows (see appendix),
\be
 |\Phi|^2={2i\over \S-\bar \S}.
\ee


\subsection{Supersymmetric case}

Let us first review how the above framework can be used to describe the effective field theory of an exactly $\N=4$ supersymmetric heterotic model in Minkowski space~\cite{GP1,GP2}. Some of the quantum numbers characterizing the spectrum are generalized momenta $p=(p_L,p_R)$, which take values in a Narain lattice $\Lambda_{(6,22)}$. The latter is  a moduli-dependent, even, self-dual and Lorentzian lattice of signature $(6,22)$~\cite{Narain1,Narain2}. There are 22 generically massless vector multiplets satisfying $p=0$, which realize the $U(1)^{22}$ Cartan sub-algebra of the gauge group generated by the right-moving bosonic string. Other vector multiplets satisfying the level matching condition $-p_{\rm L}^2+p_{\rm R}^2=2$ become massless at enhanced symmetry points in moduli space, where their momenta satisfy $p_{\rm L}^2=p_{\rm R}^2-2=0$. These states  admit  towers of pure KK or winding modes along internal directions (for the level matching condition $-p_{\rm L}^2+p_{\rm R}^2=2$ to remain valid), which may be light (but not massless) far enough from the core of the moduli space. As a result, the low energy effective supergravity valid everywhere in moduli space should take all of these vector multiplets into account, implying $k$ to be infinite~\cite{GP1,GP2}. In that case, the off-shell action would be invariant under the full T-duality group $O(6,22,\Z)$~\cite{GP1,GP2,T-dualityGPR,Polchinski_book2}. However, in any finite region in moduli space, only a finite number of  vector multiplets satisfying $-p_{\rm L}^2+p_{\rm R}^2=2$ are lighter than some given cutoff scale. In such a domain, the effective supergravity may be restricted to a finite  number $k$ of light vector multiplets, with all others integrated out. 

Whether $k$ is finite or not, it is convenient to define the index $S$ to take the values $1,\dots,6$ associated with the Abelian generators of the $U(1)^6$ gauge symmetry generated by the graviphotons, or $7,\dots, 28$ for the $U(1)^{22}$ Cartan sub-algebra, or finally any  ``generalized root $p\in \Lambda_{(6,22)}$ of squared length equal to 2'' that yields a light vector multiplet in the moduli space region under interest.  The use of the word ``root'' is justified by the fact that in a Cartan-Weyl basis, the components of $p=(p_{\rm L},p_{\rm R})$ are the charges of the associated vector multiplet under $U(1)^6\times U(1)^{22}$~\cite{GP1,GP2}. Therefore, the $p$'s are the data needed to describe the gauge interactions i.e. the structure constants $f_{RST}$. Notice however that to write the off-shell effective supergravity valid in a given region in moduli space, we may choose any background value of the moduli fields in this region to compute the momenta~$p$~\cite{AK}.  Consistently, the effective field theory must be independent of this choice.  To be specific, up to permutations of the indices, the non-vanishing structure constants are~\cite{GP1,GP2} 
\be
\begin{aligned}
f_{Spp'} &= \langle p_{S+3, \rm L}\rangle\, \delta_{p+p',0},\quad S\in\{1,\dots,6\},\\
f_{Spp'} &= \langle p_{S-3, \rm R}\rangle \, \delta_{p+p',0},\quad S\in\{7,\dots,28\},\\
f_{pp'p^{\prime\prime}}&=\varepsilon(p,p')\, \delta_{p+p'+p^{\prime\prime},0}, \quad \where \quad -p_{\rm L}^2+p_{\rm R}^2=-p_{\rm L}^{\prime 2}+p_{\rm R}^{\prime 2}=-p^{\prime\prime 2}_{\rm R}+p^{\prime\prime 2}_{\rm L}=2,
\end{aligned}
\label{gauging}
\ee
while $\varepsilon(p,p')$ are suitable signs, and brackets $\langle\,  \cdots\rangle$ stand for background values. 

To set these ideas on a simple example, we can consider the $\N=4$ model obtained by compactifying on the background~(\ref{back}), when no Scherk-Schwarz mechanism is implemented. The associated partition function is given in Eq.~(\ref{mod}), with the cocycle $\C$ omitted. The states $m_4=-n_4=\epsilon$, $m_5=n_5=\dots=m_9=n_9=0$, where $\epsilon=\pm 1$, are massless at the enhanced $SU(2)$ symmetry point $R_4=1$. However, when the $T^2\times T^2$ moduli are of order 1 and generic, the KK or winding modes
\be
\begin{aligned}
&p_{4\underset{\,\scriptstyle \rm R}{\rm L}}={\epsilon\over \sqrt{2}} \Big({1\over R_4}\mp R_4\Big), \quad p_{5\rm L}=p_{5\rm R}={m_5\over \sqrt{2} R_5}\neq 0, \;\;\:\quad p_{6\underset{\,\scriptstyle \rm R}{\rm L}}=\dots=p_{9\underset{\,\scriptstyle \rm R}{\rm L}}=0,\\
\and \quad &p_{4\underset{\,\scriptstyle \rm R}{\rm L}}={\epsilon\over \sqrt{2}} \Big({1\over R_4}\mp R_4\Big), \quad p_{5\rm L}=-p_{5\rm R}={n_5\over \sqrt{2}} R_5\neq 0, \quad p_{6\underset{\,\scriptstyle \rm R}{\rm L}}=\dots=p_{9\underset{\,\scriptstyle \rm R}{\rm L}}=0,
\end{aligned}
\label{states45}
\ee
are  respectively light for large enough $R_5$, or low enough $R_5$, when $R_4$ sits in  the vicinity of 1. Therefore, an effective description valid  in the region $R_4\simeq 1$ for arbitrary $R_5$ can be constructed by including both towers of vector multiplets. Of course, when $R_5\gg 1$, the degrees of freedom wrapped  along $S^1(R_5)$ are very heavy and must be set to 0, while for $R_5\ll 1$ it is the KK states propagating along $S^1(R_5)$ that must be frozen at their trivial background values. For $R_5\simeq 1$, the modes $m_5=n_5=0$ are the only ones dynamical. In that case, the non-vanishing structure constants to be considered for an effective description valid for arbitrary $R_5$ are~\cite{AK}
\be
f_{1,p,-p}=\langle p_{4\rm L}\rangle, \quad f_{2,p,-p}=\langle p_{5\rm L}\rangle,\quad f_{7,p,-p}=\langle p_{4\rm R}\rangle, \quad f_{8,p,-p}=\langle p_{5\rm R}\rangle,
\label{fpp}
\ee
where we set $\epsilon =1$ in the expressions of $p_{4\rm L}, p_{4\rm R}$.
If the effective action is independent of the choice of background $\langle R_4\rangle$ around 1,  taking $\langle R_4\rangle=1$ (with $\langle R_5\rangle$ arbitrary) is particular in the sense that the $SU(2)$ structure constants become explicit, since $\langle p_{4\rm R}\rangle = \epsilon \sqrt{2}$ are the $SU(2)$ roots, and  $\langle p_{4\rm L}\rangle = 0$.

In general, when a $\Z_2\times \Z_2$ modding  reduces $\N=4$ to $\N=1$ in a model, the $2+6k$ real scalars which live on the product manifold~(\ref{6k}) are reduced to $1+k_1+k_2+k_3$ complex scalars in the untwisted sector. They parameterize the descendent untwisted moduli space~\cite{truncation}
\be
{SU(1,1)\over U(1)}\times {SO(2,k_1)\over SO(2)\times SO(k_1)}\times {SO(2,k_2)\over SO(2)\times SO(k_2)}\times {SO(2,k_3)\over SO(2)\times SO(k_3)},
\label{4cosets}
\ee
whose factors are  \Ka manifolds. These complex scalars are associated with $1+k_1+k_2+k_3$ chiral multiplets, while the spectrum also contains $k-k_1-k_2-k_3$  $\N=1$ vector multiplets. Denoting respectively $K_{(0)}$, $K_{(1)}$, $K_{(2)}$, $K_{(3)}$  the \Ka potentials of the above scalar manifolds, the gravitino mass and potential in Einstein frame and $\N=1$ supergravity language satisfy~\cite{N=1-1,N=1-2} 
\be
m_{{3\over 2}\rm E}^2=e^K|W|^2,\qquad V=e^K|W|^2\left[\Big(K_i+{W_{i}\over W}\Big)K^{i\bj}\Big(K_{\bj}+{\overline{W}_{\!\bj}\over \overline{W}}\Big)-3\right]\!, 
\label{mV}
\ee 
where $K=K_{(0)}+K_{(1)}+K_{(2)}+K_{(3)}$, $W$ is the superpotential,  and subscripts $i$ or $\bi$ stand for holomorphic or antiholomorphic derivatives with respect to the $i$-th scalar fields. Identifying $m_{{3\over 2}\rm E}$ with the mass of the surviving combination of gravitini of the  $\N=4$ parent supergravity, one shows that $W$ only involves structure constants $f_{RST}$ having one index in each of the last three cosets in~(\ref{4cosets})~\cite{truncation}, namely
\be
\begin{aligned}
f_{RST}, \quad &R\in\{1,2,6+1,\dots, 6+k_1\}, \\
&S\in\{3,4,6+k_1+1,\dots, 6+k_1+k_2\},\\
&T\in\{5,6,6+k_1+k_2+1,\dots, 6+k_1+k_2+k_3\}.
\end{aligned}
\ee
Note that this result is valid whether the gauging induces or not a super-Higgs mechanism. 
In the present case, where $\N=4$ supersymmetry is exact, the non-trivial structure constants in Eq.~(\ref{fpp}) have one index $R\in\{1,2,7,8\}$, i.e. in the second coset of~(\ref{4cosets}). Therefore,  the scalar degrees of freedom labelled $p$ and $-p$ must sit, say,  in the third and fourth cosets, respectively. In that case, $W$ and thus $V$ do not vanish identically, which allows the scalars with quantum numbers given in Eq.~(\ref{states45}) to have non-trivial masses in the supergravity description, as $R_4$ and $R_5$ vary~\cite{AK}. Moreover, because the potential we are interested in involves scalars arising from the untwisted sector only, it can either be computed by using the $\N=1$ formula in Eq.~(\ref{mV}), or the $\N=4$ result given in Eq.~(\ref{V4}).\footnote{Consistently, we have checked that both expressions yield identical results when the states $p$ in Eq.~(\ref{states45}) are restricted to $m_5=n_5=0$.} Finally, due to the orbifold action, the complex scalars with quantum numbers  $p$ and $-p$  have to be identified. 


\subsection{Non-supersymmetric case}

In the background~(\ref{back}) modded by $\Z_2\times \Z_2$, the spontaneous breaking of $\N=1$ supersymmetry we consider is realized by coupling  the lattice of zero modes associated with $S^1(R_4)$ to the boundary conditions of the worldsheet fermions $\psi^6,\psi^8$. For simplicity, we will describe the potential of the effective supergravity for a minimal set of  degrees of freedom. We restrict to the dilaton $\phi_{\rm dil}$, the radii $R_4,R_5$, and the potentially tachyonic real scalars described at the end of Sect.~\ref{secZ}, which have vanishing momenta and winding numbers along $T^2\times T^2$. However, our analysis could be generalized to include more moduli and potentially tachyonic modes.  
The relevant  $\sigma$-model to start with is based on the target space
 \be
{SU(1,1)\over U(1)}\times {SO(2,2)\over SO(2)\times SO(2)}\times {SO(2,k_+)\over SO(2)\times SO(k_+)}\times {SO(2,k_-)\over SO(2)\times SO(k_-)},\label{cosets+-}
\ee
which deserves comments. The coordinates of the second coset are associated with the metric and antisymmetric tensor moduli fields of the internal 2-torus of coordinates $X^4,X^5$, which we will actually take in factorized form. The third coset is parameterized by the potentially tachyonic real scalars whose quantum numbers are 
\be
\begin{aligned}
&\,\!\!\!\!\!|2m_4+1=-n_4=+1,m_5=0,n_5=0\rangle, \\
&\,\!\!\!\!\!|2m_4+1=-n_4=+1,m_5,n_5=0\rangle, \quad m_5\neq 0,\\
&\,\!\!\!\!\!|2m_4+1=-n_4=+1,m_5=0,n_5\rangle, \quad n_5\neq 0,
\end{aligned}
\label{se+}
\ee 
together with an equal number of real scalar superpartners that will be massive once the gauging is implemented. Together, they  realize the bosonic parts of $k_+=+\infty$ chiral multiplets. Similarly, the coordinates on the fourth coset are associated with the potentially tachyonic real degrees of freedom  
\be
\begin{aligned}
&|2m_4+1=-n_4=-1,m_5=0,n_5=0\rangle, \\
&|2m_4+1=-n_4=-1,-m_5,n_5=0\rangle, \quad m_5\neq 0,\\
&|2m_4+1=-n_4=-1,m_5=0,-n_5\rangle, \quad n_5\neq 0,
\end{aligned}
\label{se-}
\ee
together with an equal number of real superpartners that will be massive after gauging. They realize the bosonic parts of $k_-=+\infty$ chiral multiplets. Due to the action of $\boldsymbol{G_2}$ (see Eq.~(\ref{rrrr})), the states~(\ref{se+}) and~(\ref{se-}) for given $m_5, n_5$ have to be identified. 

As reviewed in the appendix, the coset  ${SO(2,2)\over SO(2)\times SO(2)}$ can be parameterized by constrained fields $\phi^1, \phi^2,\phi^7,\phi^8$ that can be expressed in terms of 2 complex variables $\T,\U$, 
\be
\begin{aligned}
&\phi^1={1-\T\U\over \sqrt{Y_{(1)}}}, \quad \phi^2={\T+\U\over \sqrt{Y_{(1)}}},\quad \phi^7={1+\T\U\over \sqrt{Y_{(1)}}}, \quad \phi^8={\T-\U\over \sqrt{Y_{(1)}}},\\
&\where \quad Y_{(1)}=-(\T-\bar \T)(\U-\bar \U)>0.\esp
\end{aligned}
\label{TU}
\ee 
In order to define coordinates of the manifolds ${SO(2,k_+)\over SO(2)\times SO(k_+)}$ and ${SO(2,k_-)\over SO(2)\times SO(k_-)}$, it is convenient to introduce indices $A$ and $\tilde A$ that label the states~(\ref{se+}) and~(\ref{se-}), respectively. Denoting shortly
\be
\begin{aligned}
A&=(+,0,0),\; \,(+,m_5,0) \;\mbox{ or } \;(+,0,n_5)\\
\tilde A&=(-,0,0),\; \,(-,-m_5,0)\; \mbox{ or } \;(-,0,-n_5), \quad \where \quad m_5,n_5\neq 0,
\end{aligned}
\ee
the cosets are parameterized by the constrained fields $\phi^3,\phi^4,\phi^A$ and $\phi^5,\phi^6,\phi^{\tilde A}$ that depend on unconstrained variables $\omega_A$ and $\omega_{\tilde A}$ to be associated with the modes~(\ref{se+}) and~(\ref{se-}). Using Eq.~(\ref{soCV}), we have 
\be
\begin{aligned}
&\qquad \qquad \phi^3={1\over 2\sqrt{Y_{(+)}}}\Big(1+\sum_B\omega_B\Big),\qquad\qquad  \phi^5={1\over 2\sqrt{Y_{(-)}}}\Big(1+\sum_{\tilde B}\omega_{\tilde B}\Big),\\
&\qquad \qquad\phi^4={i\over 2\sqrt{Y_{(+)}}}\Big(1-\sum_B\omega_B\Big),\qquad \qquad\phi^6={i\over 2\sqrt{Y_{(-)}}}\Big(1-\sum_{\tilde B}\omega_{\tilde B}\Big),\\
&\qquad \qquad\!\phi^{A}={\omega_A\over \sqrt{Y_{(+)}}},\qquad \qquad \qquad\qquad \qquad\,\phi^{\tilde A}={\omega_{\tilde A}\over \sqrt{Y_{(-)}}},\\ 
&\;\;\;\quad \qquad \where \quad Y_{(+)}=1-2\sum_B|\omega_B|^2+\Big|\sum_B\omega_B^2\Big|^2>0,\\
& \qquad  \qquad  \qquad  \quad\;\, Y_{(-)}=1-2\sum_{\tilde B} |\omega_{\tilde B} |^2+\Big|\sum_{\tilde B}\omega_{\tilde B}^2\Big|^2>0.
\end{aligned}
\label{om}
\ee

The relations~(\ref{gauging}) between the constants $f_{RST}$ of the $\N=4$ gauging and the quantum numbers $p=(p_{\rm L},p_{\rm R})$ have been derived in the supersymmetric case in Refs~\cite{GP1,GP2}. As far as we know, the generalization of these results in presence of super-Higgs effect responsible for the $\N=4\to \N=0$ spontaneous breaking of supersymmetry is not known.\footnote{The cases restricted to the untwisted sectors of exact $\N=2$ and $\N=1$ models obtained by orbifold actions on parent $\N=4$ theories have also been treated in Refs~\cite{GP1,GP2}.} The main novelty is that $p_{\rm L},p_{\rm R}$ do not take anymore universal values among the 8 bosonic and 8 fermionic degrees of freedom of the $\N=4$ vector multiplets. To be specific, let us denote $Q$ and $Q_2,Q_3,Q_4$ the 4 Cartan charges of  the $SO(8)$ little group generated by the left-moving supersymmetric side of the heterotic worldsheet. We have $Q=-N+{\alpha\over 2}$ and, as seen in Eqs~(\ref{Nq}),~(\ref{pq}), the left- and right-moving quantum charges in presence of deformation ($e=1$) of the $\psi^6,\psi^8$ boundary conditions become
\begin{align}
p_{4\rm L}&={1\over \sqrt{2}}\Big({m_4+eQ-{1\over 2}n_4e^2 \over R_4}+ n_4R_4\Big), \;\; Q'=Q-n_4e, \;\;Q'_2=Q_2, \;\;Q'_3=Q_3, \;\;Q'_4=Q_4,\nonumber \\
p_{4\rm R}&={1\over \sqrt{2}}\Big({m_4+eQ-{1\over 2}n_4e^2 \over R_4}-n_4R_4\Big).
\end{align}
They mix non-trivially $m_4,n_4$ with  $\vec Q=(Q,Q_2,Q_3,Q_4)$, so that $\Lambda_{(6,22)}$ should be extended to a larger Lorentzian lattice, whose vectors $(p_{\rm L},\vec Q',p_{\rm R})$ have norm $-p_{\rm L}^2-\vec Q^{\prime 2}+p_{\rm R}^2$. However, in our parent model of interest, the  $\N=4\to \N=0$ spontaneous breaking implies all scalar superpartners of the states $A$ and $\tilde A$ to be massive. After implementation of the $\Z_2\times \Z_2$ orbifold action, restricting the dynamics to the potentially tachyonic modes only, the only relevant Cartan charges $p_{\rm L},p_{\rm R}$ are  therefore those associated  with  the potentially tachyonic real scalars $A$ and $\tilde A$, which are  identified.  Hence, we switch on structure constants 
\begin{align}
&f_{1A{\tilde A}}=\langle p_{4{\rm L}A}\rangle\equiv \langle p_{4{\rm L}}\rangle={1\over\sqrt{2}}\Big({1\over 2 \langle R_4\rangle}-\langle R_4\rangle\Big), &&\!\!f_{2A{\tilde A}}=\langle p_{5{\rm L}A}\rangle=0,  \;{m_5\over \sqrt{2}\, \langle R_5\rangle}\;\, \mbox{or}\,\; {n_5\over \sqrt{2}}\, \langle R_5\rangle\nonumber ,\espD\\
&f_{7A{\tilde A}}=\langle p_{4{\rm R}A}\rangle\equiv \langle p_{4{\rm R}}\rangle={1\over\sqrt{2}}\Big({1\over 2\langle R_4\rangle}+\langle R_4\rangle\Big),&&\!\! f_{8A{\tilde A}}=\langle p_{5{\rm R}A}\rangle=0,  \;{m_5\over \sqrt{2}\, \langle R_5\rangle}\;\, \mbox{or}\,\; -{n_5\over \sqrt{2}}\, \langle R_5\rangle,\nonumber\espD \\
&&&\!\!\!\!\!\!\!\!\!\!\!\!\!\!\!\!\!\!\!\!\!\!\!\!\!\!\!\!\!\!\!\!\!\!\!\!\!\!\!\!\!\!\!\!\!\!\!\! \!\!\!\!\!\!\!\!\!\!\!\!\!\!\!\!\!\!\!\!\!\!\!\!\!\!\!\!\!\!\!\!\!\!\!\!\!\!\!\!\!\!\!\!\!\!\!\!\!\!\!\!\!\!\!\!\!\!\!\!\where \quad A=(+,0,0),\;\, (+,m_5,0) \;\mbox{ or }\; (+,0,n_5).\esp
\label{fiib}
\end{align}
Notice that the level matching condition now reads $-\langle p_{\rm L}\rangle^2+\langle p_{\rm R}\rangle^2=1$. 
In the notations of the above formula, it is understood that $A$ and ${\tilde A}$ are not independent indices in the sense that states $A$ and ${\tilde A}$ always have opposite momenta and winding numbers. 
However, the structure constants~(\ref{fiib}) being formally similar to those encountered in the supersymmetric case, Eq.~(\ref{fpp}), they cannot induce any super-Higgs mechanism. In particular, the real and imaginary parts of the $\omega_A$'s and $\omega_{{\tilde A}}$'s would be treated on equal footing and have degenerate masses. 

In order to break spontaneously $\N=4\to \N=0$ in the parent supergravity, a non-Abelian structure among the 6 graviphotons must be implemented. However, the generated potential should admit a phase compatible with  Minkowski space-time. This has to be the case since the underlying string theory is a no-scale model~\cite{noscale}, provided $R_4$ sits outside the range~(\ref{range}). By definition, structure constants satisfying these conditions define the so-called ``flat gaugings''~\cite{SS2}. After $\Z_2\times \Z_2$ truncation, in $\N=1$ supergravity language, for the superpotential $W$ and thus the potential $V$ to be affected by the non-Abelian interactions of the graviphotons in the parent theory, additional structure constant $f_{RST}$ having one index in each of the last three cosets in~(\ref{4cosets}) must be considered~\cite{AK,ADK}. Because our choice of implementation of the stringy Scherk-Schwarz mechanism involves the left-and right-moving quantum numbers along $S^1(R_4)$ only, we switch on  
\be
\begin{aligned}
f_{135}=e_{\rm L},\quad \;\;f_{735}=e_{\rm R},\quad \;\;f_{146}=\tilde e_{\rm L},\;\;\quad f_{746}=\tilde e_{\rm R},
\end{aligned}
\label{ee}
\ee
where $e_{\rm L},e_{\rm R}$ and $\tilde e_{\rm L},\tilde e_{\rm R}$ have to be determined for the mass spectrum in the no-scale supergravity phase to match that of the underlying string model. Notice that if only  $e_{\rm L},e_{\rm R}$ have been considered in Refs~\cite{AK,ADK}, we will see that $\tilde e_{\rm L},\tilde e_{\rm R}$ play an important role for matching an enlarged spectrum. 


\subsection{Effective potential in the non-supersymmetric case}

In order to write the potential $V$ that involves only untwisted states of the $\N=1\to \N=0$ orbifold theory, we find easier to use the $\N=4$ supergravity expression~(\ref{V4}) rather than its $\N=1$ counterpart, Eq.~(\ref{mV}). The link between the constrained fields $\phi^S$ in Eqs~(\ref{TU}) and~(\ref{om}) and the $\N=4$  variables $Z^S_a$ are provided by the relations (see Eq.~(\ref{phiZ}))
\be
\begin{aligned}
&\phi^S={1\over 2}\big( Z^S_1+iZ^S_2\big),\quad \mbox{for }\;\; S=1,2,7,8,\\
&\phi^S={1\over 2}\big(Z^S_3+iZ^S_4\big),\quad \mbox{for }\;\; S=3,4,A,\\
&\phi^S={1\over 2}\big(Z^S_5+iZ^S_6\big),\quad \mbox{for }\;\; S=5,6,\tilde A,\\
\end{aligned}
\ee
with all other $Z^S_a\equiv 0$. In that case, Eq.~(\ref{var}) reduces consistently to 
\be
\begin{aligned}
&\hat \eta_{ST} Z^S_aZ^T_b=-\delta_{ab}, \quad a,b\in\{1,2\},\\
&\check\eta_{ST} Z^S_aZ^T_b=-\delta_{ab}, \quad a,b\in\{3,4\},\\
&\check \eta_{ST} Z^S_aZ^T_b=-\delta_{ab}, \quad a,b\in\{5,6\},\\
\end{aligned}
\ee 
where $\hat \eta=\mbox{diag}(-1,-1,1,1)$ and $\check \eta=\mbox{diag}(-1,-1,1,\dots)$. As a result, the non trivial real scalars $Z^{ST}$ appearing in Eq.~(\ref{V4}) have both indices $S,T$ in either of the last 3 cosets~(\ref{cosets+-}), namely 
\be
Z^{ST}=2\big(\phi^S\bar \phi^T+\bar \phi^S\phi^T\big),  \quad \mbox{for}\quad S,T=1,2,7,8\quad \mbox{or} \quad 3,4,A\quad \mbox{or}\quad  5,6,\tilde A.
\label{zST}
\ee
It turns out that the computation of $V$ is greatly simplified by the introduction of indices correlated as follows:
\be
\mbox{For}\quad  \A=3,\;\, 4\;\mbox{ or }\, A, \quad \mbox{we define}\quad  \tilde \A=5,\;\, 6\;\mbox{ or } \,{\tilde A},\quad \mbox{respectively}.
\ee
With this convention, the structure constants with one index equal to 1 or 7 can be unified in a single notation, 
\be
\begin{aligned}
f_{1\A\tilde \A}=\langle p_{4\rm L\A}\rangle ,\qquad f_{7\A\tilde \A}=\langle p_{4\rm R\A}\rangle,
\end{aligned}
\ee
which amounts to writing $\langle p_{4\rm \underset{\scriptstyle \rm R}{\rm L}3}\rangle\equiv e_{\underset{\scriptstyle \rm R}{\rm L}}$ and $\langle p_{4\rm \underset{\scriptstyle \rm R}{\rm L}4}\rangle\equiv \tilde e_{\underset{\scriptstyle \rm R}{\rm L}}$. In total, the potential takes the form 
\be
\begin{aligned}
V={|\Phi|^2\over 2}\bigg\{&\sum_{\A,\B} \Big[\Big(\eta^{\A\B}\big(Z^{\A\B}+Z^{\tilde \A\tilde \B})+2Z^{\A\B}Z^{\tilde \A\tilde \B}\big)\Big)v_{44}(\T,\U,\langle p_{4\A}\rangle,\langle p_{4\B}\rangle)\\
&\;\;\quad +Z^{\A\B}Z^{\tilde \A\tilde \B}\big(p_{4{\rm R}\A}\, p_{4{\rm R}\B}-p_{4{\rm L}\A}\, p_{4{\rm L}\B}\big)\Big]\espD\\
&\!\!\!\!\!\!+\sum_{A,B} \Big[\Big(\eta^{AB}\big(Z^{AB}+Z^{\tilde A\tilde B}\big)+2Z^{AB}Z^{\tilde A\tilde B}\Big)v_{55}(\T,\U,\langle p_{5A}\rangle,\langle p_{5B}\rangle)\\
&\;\;\quad +Z^{AB}Z^{\tilde A\tilde B}\big(p_{5{\rm R}A}\, p_{5{\rm R}B}-p_{5{\rm L}A}\, p_{5{\rm L}B}\big)\Big]\espD\\
&\!\!\!\!\!\! +\sum_{\A,B} \Big(\eta^{\A B}\big(Z^{\A B}+Z^{\tilde \A\tilde B}\big)+2Z^{\A B}Z^{\tilde \A\tilde B}\Big)2v_{45}(\T,\U,\langle p_{4\A}\rangle,\langle p_{5B}\rangle)\bigg\},
\end{aligned}
\ee
where the sums are indicated explicitly for clarity. In the above formula, we have defined $p_{I\A}\equiv(p_{I{\rm L}\A},p_{I{\rm R}\A})$, $I\in\{4,5\}$, and 
\be
\begin{aligned}
v_{44}(\T,\U,x,y)&=Z^{11}x_{\rm L}y_{\rm L}+Z^{17}(x_{\rm L}y_{\rm R}+y_{\rm L}x_{\rm R})+Z^{77}x_{\rm R}y_{\rm R},\\
v_{55}(\T,\U,x,y)&=Z^{22}x_{\rm L}y_{\rm L}+Z^{28}(x_{\rm L}y_{\rm R}+y_{\rm L}x_{\rm R})+Z^{88}x_{\rm R}y_{\rm R},\\
v_{45}(\T,\U,x,y)&=Z^{12}x_{\rm L}y_{\rm L}+Z^{18}x_{\rm L}y_{\rm R}+Z^{72}y_{\rm L}x_{\rm R}+Z^{78}x_{\rm R}y_{\rm R}.
\end{aligned}
\ee
In the present work, because we impose the first internal 2-torus to be $S^1(R_4)\times S^1(R_5)$, we can restrict the moduli $\T$ and $\U$ to be purely imaginary. We define
\be
\T=i\, \cR_4\cR_5, \qquad \U=i\, {\cR_4\over \cR_5},
\ee
where the precise relation between the real variables $\cR_4,\cR_5$ and the worldsheet moduli $R_4,R_5$ will have to be determined. From the relation~(\ref{zST}) for $S,T\in\{1,2,7,8\}$, we obtain~
\be
v_{44}(\T,\U,x,y) \equiv  v(\cR_4,x,y), \quad v_{55}(\T,\U,x,y) \equiv  v(\cR_5,x,y),\quad v_{45}(\T,\U,x,y) \equiv  0, 
\ee
where we have defined 
\be
\begin{aligned}
v(\cR,x,y)&= {1\over 4}\Big[\Big({1\over \cR}+\cR\Big)x_{\rm L}+\Big({1\over \cR}-\cR\Big)x_{\rm R}\Big]\Big[\Big({1\over \cR}+\cR\Big)y_{\rm L}+\Big({1\over \cR}-\cR\Big)y_{\rm R}\Big]\espD\\
&={1\over 4}\Big[{(x_{\rm L}+x_{\rm R})(y_{\rm L}+y_{\rm R})\over \cR^2}+2(x_{\rm L}y_{\rm L}-x_{\rm R}y_{\rm R})+\cR^2(x_{\rm L}-x_{\rm R})(y_{\rm L}-y_{\rm R})\Big].
\end{aligned}
\ee
The expressions of the $Z^{\A\B}$'s and $Z^{\tilde \A\tilde \B}$'s involve 
\be
\Omega_A={\omega_A\over \sqrt{Y_{(+)}}},\qquad \Omega_{\tilde A}={\omega_{\tilde A}\over \sqrt{Y_{(-)}}},
\ee 
in terms of which we have 
\be
\begin{aligned}
&Z^{33}=1+\sum_A (\Omega_A+\bar \Omega_A)^2,&&Z^{55}=1+\sum_{\tilde A} (\Omega_{\tilde A}+\bar \Omega_{\tilde A})^2, \espD \\
&Z^{44}=1-\sum_A (\Omega_A-\bar \Omega_A)^2&&Z^{66}=1-\sum_{\tilde A} (\Omega_{\tilde A}-\bar \Omega_{\tilde A})^2,\espD\\
&Z^{AB}= 2(\Omega_A\bar\Omega_B+\bar \Omega_A\Omega_B),&&Z^{\tilde A\tilde B}= 2(\Omega_{\tilde A}\bar\Omega_{\tilde B}+\bar \Omega_{\tilde A}\Omega_{\tilde B}),\espD\\
&Z^{34}=-i \sum_A(\Omega_A^2-\bar \Omega_A^2),&& Z^{56}=-i \sum_{\tilde A}(\Omega_{\tilde A}^2-\bar \Omega_{\tilde A}^2),\espD\\
&Z^{3A}=-\Omega_A{1+\sum_B\bar \omega_B^2\over \sqrt{Y_{(+)}}}+\cc,&&Z^{5\tilde A}=-\Omega_{\tilde A}{1+\sum_{\tilde B}\bar \omega_{\tilde B}^2\over \sqrt{Y_{(-)}}}+\cc,\espD \\
&Z^{4A}=i\Omega_A{1-\sum_B\bar \omega_B^2\over \sqrt{Y_{(+)}}}+\cc,&&Z^{6\tilde A}=i\Omega_{\tilde A}{1-\sum_{\tilde B}\bar \omega_{\tilde B}^2\over \sqrt{Y_{(-)}}}+\cc.
\end{aligned}
\ee

To proceed, we make some remarks on the expansion in $\omega_A,\omega_{\tilde A}$ of the potential:

\noindent $\bullet$ At zeroth order, i.e. with no tachyon  condensation, only $Z^{33},Z^{55}$ and $Z^{44},Z^{66}$ are non-trivial. As a result, up to the overall dressing by $|\Phi|^2$, $V$ reduces to a constant expressed in terms of $e_{\rm L}, e_{\rm R}$ and  $\tilde e_{\rm L}, \tilde e_{\rm R}$. Since this configuration should describe the no-scale phase characterized by a vanishing cosmological constant, this constant must vanish. 

\noindent $\bullet$ At next order, $V$ contains quadratic terms in $\Re\omega_A,\Re\omega_{\tilde A}$, which depend on $e_{\rm L},e_{\rm R}$ but not in $\tilde e_{\rm L},\tilde e_{\rm R}$. $V$ also contains quadratic terms in $\Im\omega_A,\Im\omega_{\tilde A}$, which depend on $\tilde e_{\rm L},\tilde e_{\rm R}$ but not in $e_{\rm L},e_{\rm R}$. Hence, it is a matter of convention to choose $e_{\rm L},e_{\rm R}$ rather than $\tilde e_{\rm L},\tilde e_{\rm R}$ to reproduce the tachyonic mass terms (in Einstein frame) of the underlying string model. In that case, $\Re\omega_A,\Re\omega_{\tilde A}$ are the associated degrees of freedom and $\Im\omega_A,\Im\omega_{\tilde A}$ are massive superpartners to be set to 0. $\tilde e_{\rm L},\tilde e_{\rm R}$ can then be tuned to satisfy the above mentioned cosmological constant constraint. In fact, for these statements to be true, the kinetic terms should also be block-diagonal in $\Re\omega_A,\Re\omega_{\tilde A}$, and in $\Im\omega_A,\Im\omega_{\tilde A}$. This turns out to be the case,  since the scalar kinetic terms of the truncated supergravity are determined by the \Ka metric~\cite{truncation}, and  take the following form at quadratic order,
\be
\begin{aligned}
-g^{\mu\nu}\Big(K_{\omega_A\bar \omega_A}&\, \partial_\mu \omega_A\partial_\nu\bar \omega_{A}+K_{\omega_{\tilde A}\bar\omega_{\tilde A}}\,\partial_\mu \omega_{\tilde A}\partial_\nu\bar \omega_{\tilde  A}\Big)\\
&=2g^{\mu\nu}\Big(\partial_\mu(\Re \omega_A)\partial_\nu(\Re \omega_A)+\partial_\mu(\Re \omega_{\tilde A})\partial_\nu(\Re \omega_{\tilde A})\\
&\qquad \;\:\,+\partial_\mu(\Im \omega_A)\partial_\nu(\Im \omega_A)+\partial_\mu(\Im \omega_{\tilde A})\partial_\nu(\Im \omega_{\tilde A})+\O(\partial_\mu\partial_\nu\omega^4)\Big).
\end{aligned}
\label{kinetic terms}
\ee
From now on, we thus take
\be
\omega_A, \omega_{\tilde A}\in\R\quad \Longrightarrow\quad \Omega_A={\omega_A\over 1-\sum_B\omega_B^2}, \quad \Omega_{\tilde A}={\omega_{\tilde A}\over 1-\sum_{\tilde B}\omega_{\tilde B}^2},
\ee
which yields
\be
\begin{aligned}
&Z^{33}=1+4\sum_A \Omega_A^2,&&Z^{55}=1+4\sum_{\tilde A} \Omega_{\tilde A}^2,  \\
&Z^{44}=1,&&Z^{66}=1,\\
&Z^{AB}= 4\Omega_A\Omega_B,&&Z^{\tilde A\tilde B}= 4\Omega_{\tilde A}\Omega_{\tilde B},\\
&Z^{34}=0,&& Z^{56}=0,\\
&Z^{3A}=-2\Omega_A\Big({1+4\sum_B\Omega_B^2}\Big)^{1\over 2},&&Z^{5\tilde A}=-2\Omega_{\tilde A}\Big(1+4\sum_{\tilde B}\Omega_{\tilde B}^2\Big)^{1\over 2}, \\
&Z^{4A}=0,&&Z^{6\tilde A}=0.
\end{aligned}
\ee

In the present work, we consider a deformation of the boundary conditions of the worldsheet fermions $\psi^6,\psi^8$, which leads to the potentially tachyonic spectrum described at the end of Sect.~\ref{secZ}. This amounts to identifying suitably the degrees of freedom $\omega_A$ and $ \pm\omega_{\tilde A}$, where the choice of sign turns out to be a matter of convention. This follows from the fact that the transformation $\omega_{\tilde A}\to -\omega_{\tilde A}$ is equivalent to $Z^{5\tilde A}\to -Z^{5\tilde A}$, and that the latter can be compensated by a flip $(e_{\rm L},e_{\rm R})\to -(e_{\rm L},e_{\rm R})$. Making the choice 
\be
\omega_A\equiv \omega_{\tilde A}\quad \Longrightarrow\quad \Omega_A\equiv \Omega_{\tilde A},
\ee  
the potential takes the following form,
\be
V={|\Phi|^2\over 2}\Big(C^{(0)}+C^{(2)}_A\Omega_A^2+C^{(4)}_{AB}\Omega_A^2\Omega_B^2\Big),
\ee
with constant coefficients defined as
\begin{align}
C^{(0)}&=-e_{\rm L}^2+e_{\rm R}^2-\tilde e_{\rm L}^2+\tilde e_{\rm R}^2,\nonumber\espD\\
C^{(2)}_{A}&=2\bigg[{1\over \cR_4^2}\big(\langle p_{4{\rm L}}\rangle+\langle p_{4{\rm R}}\rangle+e_{\rm L}+e_{\rm R}\big)^2+\cR_4^2\big(\langle p_{4{\rm L}}\rangle-\langle p_{4{\rm R}}\rangle+e_{\rm L}-e_{\rm R}\big)^2\espD\nonumber\\
&\quad \;\;+\!{1\over \cR_5^2}\big(\langle p_{5{\rm L}A}\rangle +\langle p_{5{\rm R}A}\rangle\big)^2+\cR_5^2\big(\langle p_{5{\rm L}A}\rangle -\langle p_{5{\rm R}A}\rangle\big)^2\nonumber\\
&\quad \;\;+2\big(\langle p_{4{\rm L}}\rangle^2-\langle p_{4{\rm R}}\rangle^2-e_{\rm L}^2+e_{\rm R}^2+\langle p_{5{\rm L}A}\rangle^2-\langle p_{5{\rm R}A}\rangle^2\big)\bigg],\espD\\
C^{(4)}_{AB}&=8\bigg[{1\over \cR_4^2}\big(\langle p_{4{\rm L}}\rangle+\langle p_{4{\rm R}}\rangle+e_{\rm L}+e_{\rm R}\big)^2+\cR_4^2\big(\langle p_{4{\rm L}}\rangle-\langle p_{4{\rm R}}\rangle+e_{\rm L}-e_{\rm R}\big)^2\espD\nonumber\\
&\quad \;\;+\!{1\over \cR_5^2}\big(\langle p_{5{\rm L}A}\rangle +\langle p_{5{\rm R}A}\rangle\big)\big(\langle p_{5{\rm L}B}\rangle +\langle p_{5{\rm R}B}\rangle\big)\!+\cR_5^2\big(\langle p_{5{\rm L}A}\rangle -\langle p_{5{\rm R}A}\rangle\big)\big(\langle p_{5{\rm L}B}\rangle -\langle p_{5{\rm R}B}\rangle\big)\bigg].\nonumber 
\end{align}
As explained before, in order to identify the structure constants responsible for the $\N=1\to \N=0$ spontaneous breaking, we use our knowledge of the cosmological constant and masses given in Eq.~(\ref{massesT}), which are valid in the no-scale supergravity phase. To work  in Einstein frame, we combine  the string theory axion $\chi$ and dilaton field $\phi_{\rm dil}$ into the axio-dilaton scalar 
\be
S=\chi+ie^{-2\phi_{\rm dil}}.
\ee
In these notations, the conditions to be imposed take the form
\be
C^{(0)}=0,\qquad {|\Phi|^2\over 2}\, C^{(2)}_A = 4 \, {2i\over S-\bar S}\, M^2_A,
\label{condis}
\ee
where the factor 4 arises from the normalization of the kinetic terms of $\Re \omega_A\equiv \Re \omega_{\tilde A}$ in Eq.~(\ref{kinetic terms}).   
However, the above masses in Einstein frame depend on  the worldsheet CFT moduli $\phi_{\rm dil}$, $R_4$, $R_5$ and the relation between them and our variables $|\Phi|^2=2i/(\S-\bar \S)$, $\cR_4$, $\cR_5$ may not be trivial. This is due to the fact that the parameterization of the 
cosets $SU(1,1)\over U(1)$ and ${SO(2,2)\over SO(2)\times SO(2)}$ in~(\ref{cosets+-}) contain some degree of arbitrariness. In particular, the \Ka potential of these manifolds being $K_{(0)}=-\ln (-i(\S-\bar \S))$ and  $K_{(1)}=- \ln Y_{(1)}$ (see Eq.~(\ref{TU})), the transformations $\S\to \gamma_S \S$ and  $(\T,\U)\to (\gamma_4\gamma_5\T,(\gamma_4/\gamma_5)\U)$ for arbitrary real constants $\gamma_S, \gamma_4,\gamma_5>0$ translate into $K_{(0)}\to K_{(0)}-\ln \gamma_S$ and $K_{(1)}\to K_{(1)}-\ln (\gamma_4\gamma_5)-\ln (\gamma_4/\gamma_5)$, which are \Ka transformations. Therefore, $\gamma_S \S$, $\gamma_4\gamma_5\T$, $(\gamma_4/\gamma_5)\U$ are as good variable as $\S$, $\T$, $\U$. In the matching of supergravity with string data,  we thus identify~
\be
\S=\gamma_S S, \quad \;\;\cR_4=\gamma_4 R_4,\quad\;\; \cR_5=\gamma_5 R_5,
\ee
with  coefficients  to be determined. 

Notice that the constraints on the masses of the modes with non-trivial momentum or winding number along $S^1(R_5)$ yield, in particular, 
\be
\begin{aligned}
&\mbox{for $A=(+,m_5,0)$ : }\quad {1\over \gamma_5^2}\big(\langle p_{5{\rm L}A}\rangle +\langle p_{5{\rm R}A}\rangle\big)^2=-{2m_5^2\over 3}\big(\langle p_{4{\rm L}}\rangle^2-\langle p_{4{\rm R}}\rangle^2-e_{\rm L}^2+e_{\rm R}^2\big),\\
&\mbox{for $A=(+,0,n_5)$ \;: }\quad\,\gamma_5^2\big(\langle p_{5{\rm L}A}\rangle -\langle p_{5{\rm R}A}\rangle\big)^2=-{2n_5^2\over 3}\big(\langle p_{4{\rm L}}\rangle^2-\langle p_{4{\rm R}}\rangle^2-e_{\rm L}^2+e_{\rm R}^2\big),
\end{aligned}
\ee
which imply
\be
{1\over \gamma_5^2\langle R_5\rangle^2}=\gamma_5^2\langle R_5\rangle^2=-{1\over 3}\big(\langle p_{4{\rm L}}\rangle^2-\langle p_{4{\rm R}}\rangle^2-e_{\rm L}^2+e_{\rm R}^2\big).
\label{nnorma}
\ee
This has several consequences. Firstly, $\gamma_5$ is related to the choice of background, $\gamma_5=1/ {\langle R_5\rangle}$. Secondly, because of the level matching condition, which fixes $-\langle p_{4{\rm L}}\rangle^2+\langle p_{4{\rm R}}\rangle^2=1$, we have $-e_{\rm L}^2+e_{\rm R}^2=-2$, and then $-\tilde e_{\rm L}^2+\tilde e_{\rm R}^2=2$ for $C^{(0)}$ to vanish. 
We stress that had we considered only the pure momentum (or pure winding) states, $\gamma_5$ would have only been related to $\langle p_{4{\rm L}}\rangle^2-\langle p_{4{\rm R}}\rangle^2-e_{\rm L}^2+e_{\rm R}^2$, leaving arbitrary  $-e_{\rm L}^2+e_{\rm R}^2=\tilde e_{\rm L}^2-\tilde e_{\rm R}^2$. This is the reason why in Refs~\cite{AK,ADK}, $\tilde e_{\rm L}, \tilde e_{\rm R}$ are not introduced (or set to 0).  It is therefore important to take into account both momentum and winding states along $S^1(R_5)$, because this fixes the r.h.s. of Eq.~(\ref{nnorma}) to 1.
In the end, the constraints on $C^{(0)}$ and $C_A^{(2)}$ admit 2 solutions, 
\be
\left\{\begin{aligned}
&e_{\rm L}=\langle p_{4\rm L}+\sigma \sqrt{3}p_{4\rm R}\rangle,\;\;  \quad e_{\rm R}=\langle p_{4\rm R}+\sigma \sqrt{3}p_{4\rm L}\rangle,\quad\;\; -\tilde e_{\rm L}^2+\tilde e_{\rm R}^2=2,\espD\\
&\gamma_S={1\over 2}, \quad \;\;\gamma_4={2+\sigma \sqrt{3}\over \langle R_4\rangle},\quad\;\; \gamma_5={1\over \langle R_5\rangle},
\end{aligned}\right.
\label{solu1}
\ee
where $\sigma=\pm 1$. In fact two more solutions exist, 
\be
\left\{\begin{aligned}
&e_{\rm L}=\sigma \sqrt{2}\, \langle p_{4\rm R}\rangle, \quad \;\;e_{\rm R}=\sigma \sqrt{2}\, \langle p_{4\rm L}\rangle,\;\;\quad -\tilde e_{\rm L}^2+\tilde e_{\rm R}^2=2,\espD\\
&\gamma_S={1\over 2}, \quad \;\;\gamma_4={\sqrt{2}+\sigma \over \langle R_4\rangle},\;\;\quad  \gamma_5={1\over \langle R_5\rangle},
\end{aligned}\right.
\ee
but it can be shown that when both the real and the  imaginary parts of the moduli $\T,\U$ are taken into account, only the solutions~({\ref{solu1}) reproduce correctly the mass spectrum arising for arbitrary metric and antisymmetric tensor backgrounds in the internal directions $X^4,X^5$. It would be interesting to see whether $\sigma$ and/or $\tilde e_{\rm L}^2=2+\tilde e_{\rm R}$ could be fixed by taking into account more degrees of freedom in the supergravity action.

We are ready to display the final expression of the low energy tree level effective potential,
\be
\begin{aligned}
V=e^{2\phi_{\rm dil}}\, 4\,&\bigg\{\Big({1\over 4R_4^2}+R_4^2-3\Big)\sum_A\Omega_A^2+\Big({1\over R_4^2}+4R_4^2\Big)\Big(\sum_A\Omega_A^2\Big)^2\\
&+ {1\over R_5^2}\sum_{m_5}m_5^2\, \Omega_{(+,m_5,0)}^2+R_5^2\sum_{n_5}n_5^2\, \Omega_{(+,0,n_5)}^2\\
& + {4\over R_5^2}\Big(\sum_{m_5}m_5\, \Omega_{(+,m_5,0)}^2\Big)^2+4R_5^2\Big(\sum_{n_5}n_5\, \Omega_{(+,0,n_5)}^2\Big)^2\bigg\}.
\end{aligned}
\label{Vfin}
\ee
In this result, we remind that the sum over $A$ runs over $(+,0,0)$, $(+,m_5,0)$ and $(+,n_5,0)$, where $m_5,n_5\neq 0$. For notational convenience, we have split the mass terms into $R_4$-dependent and $R_5$-dependent pieces. One of the greatest interest of supergravity for describing the low energy physics is that the  tree level potential captures all self-interactions in $\omega_A$'s, i.e. all $n$-points vertices, with arbitrary $n$. This very fact makes it possible to determine the allowed condensates and phases. Some remarks are in order:

\noindent  $\bullet$ The quartic terms in $\Omega_A$'s being positive, at fixed $\phi_{\rm dil}$, $V$ is bounded from below. 

\noindent $\bullet$ When $R_4$  sits outside the range~(\ref{range}), all quadratic terms in $\Omega_A$'s are positive as well. Hence, all vacua are degenerate, with vanishing cosmological constant:
\be
\langle R_4\rangle \ge R_{\rm H}\;\;\; \mbox{ or } \;\;\;\langle R_4\rangle \le R_{\rm H}, \qquad \forall A, \;\langle \Omega_A\rangle =0,\qquad \langle R_5\rangle , \langle \phi_{\rm dil}\rangle  \;\mbox{ arbitrary}.
\ee
This is the ``no-scale'' phase, where $\langle m_{{3\over 2}\rm E}\rangle \equiv \langle e^{\phi_{\rm dil}}/(2R_4)\rangle$ is arbitrary. 

\noindent $\bullet$ According to the initial string theory mass spectrum, the quadratic terms in $\omega_A$'s are invariant under the T-duality transformations $R_4\rightarrow1/(2R_4)$ and $R_5\rightarrow 1/R_5$. It turns out that the full potential $V$ and thus the full tree level bosonic action respect this T-duality.\footnote{The kinetic terms are invariant under the T-duality transformations valid in the supersymmetric case, $R_4\rightarrow1/R_4$ and $R_5\rightarrow 1/R_5$, as well as  the \Ka transformation $R_4\to 2 R_4$. Hence, they are invariant under $R_4\rightarrow1/(2R_4)$.}

\noindent $\bullet$  When $R_4$ sits in the tachyonic range~(\ref{range}),  degrees of freedom can condense. From the first line of Eq.~(\ref{Vfin}), we see that extremizing $V$ with respect to $R_4$ and $\Omega_{(+,0,0)}$  fixes $R_4=1/\sqrt{2}$ (the self-dual radius) and  the total sum $\sum_A \Omega_A^2=1/4$. To figure out which of the potentially tachyonic modes $A$ actually condense, it is enough to note that all other terms in the second and third lines of Eq.~(\ref{Vfin}) are non-negative. Hence, $V$ is minimal when all scalars with non-trivial momentum or winding number along $S^1(R_5)$ vanish. In other words, all of the condensation is supported by the tachyon  having $m_5=n_5=0$. At fixed $\phi_{\rm dil}$, there are 2 branches of minima, which are reached for the backgrounds 
\be
\begin{aligned}
&\langle \Omega_{(+,0,0)}\rangle =\pm {1\over 2},&&\quad \langle \Omega_{(+,m_5,0)}\rangle =  \langle \Omega_{(+,0,n_5)}\rangle =0, \;\; m_5,n_5\neq 0, \\
& \langle R_4\rangle ={1\over \sqrt{2}},&& \quad \langle R_5\rangle , 
\mbox{ arbitrary}.
\end{aligned}
\ee 
The $\Omega_A$'s and $R_4$ are stabilized, $R_5$ is a flat direction, and the T-duality transformation $R_4\to 1/(2R_4)$ is not spontaneously broken. The new mass spectrum can be found by expanding $V$ in small perturbations around the expectations values, 
\be
\Omega_A=\langle \Omega_A\rangle +\delta \Omega_A,\quad\;\; R_4=\langle R_4\rangle +\delta R_4,
\ee
which yields 
\be
\begin{aligned}
V=e^{2\phi_{\rm dil}}\Big( \!-1+8\delta R_4^2+16\delta\Omega_{(+,0,0)}^2\! + \frac{4}{R_5^2}\sum_{m_5}m_5^2\, \delta\Omega_{(+,m_5,0)}^2\!+4R_5^2\sum_{n_5}n_5^2\, \delta\Omega_{(+,0,n_5)}^2 + \O(\delta^3) \Big).
\end{aligned}
\ee
Strictly speaking, the word ``mass'' is a misnomer, since the dilaton has a tadpole. In fact, in terms of the string frame metric $\hat g_{\mu\nu}=e^{2\phi_{\rm dil}}g_{\mu\nu}$, the tree level action involving the Ricci scalar, dilaton and potential reads
\be
S_{\rm tree}=\int d^4x \sqrt{-\hat g}\,  e^{-2\phi_{\rm dil}}\Big({{\cal \hat R}\over 2}+2(\partial \phi_{\rm dil})^2+1+\O(\delta)\Big).
\ee
As proposed in Refs~\cite{AK,ADK}, this suggests that the condensed phase of the effective  supergravity may describe the low energy physics of a non-critical string theory, with linear dilaton background. 


\section{Conclusion}
\label{cl}

In this work, we have initiated the study of phase transitions occurring in string theory, when the scale of spontaneous breaking of supersymmetry is of the order of the string scale. Even if they are physically very different from the Hagedorn instabilities developed at high temperature, they share technical similarities about  internal or temporal cycles along which bosons and fermions have distinct boundary conditions. Significant differences nevertheless exist. 

In the Hagedorn case, 2 real scalars (in space-time dimension minus 1) become tachyonic when the radius $R_0$ of the Euclidean time circle falls below the Hagedorn radius $R_{\rm H}$. In the supersymmetry breaking case, even when the Scherk-Schwarz mechanism is implemented along a single factorized circle $S^1(R)$, the analogous 2 real scalars may be projected out of the spectrum by an orbifold action. When this arises, the ``Hagedorn-like region'' in moduli space is not the domain $R< R_{\rm H}$, but a subregion where tachyons with non-trivial momenta or winding numbers along other internal directions condense. This possibility yields interesting new phenomena that will be described elsewhere. 
Moreover, the instabilities occurring at high supersymmetry breaking scale can be analyzed when the internal metric and antisymmetric tensor are generic. In that case, target space duality transformations imply the Hagedorn-like region to be much more involved, with a fractal structure.  
 
In the present paper, we have considered a $\Z_2\times \Z_2$  heterotic orbifold setup that  illustrates the simplest situation. In this example, a real scalar with non-trivial quantum numbers only along the Scherk-Schwarz circle $S^1(R_4)$ survives the modding action, and becomes tachyonic when $R_4<R_{\rm H}$. It is accompanied by an infinite number of potentially tachyonic scalars, with momenta or winding numbers along a transverse circle $S^1(R_5)$. We have derived the tree level effective potential that depends on these degrees of freedom. It turns out to be symmetric under the T-duality transformation $R_4\to 1/(2R_4)$ and to allow only two phases.  The former is associated with the initial no-scale model, where the cosmological constant vanishes and the supersymmetry breaking scale is arbitrary. In the second phase, the tachyon with non-trivial quantum numbers only along $S^1(R_4)$ condenses, which stabilizes  $R_4$ at the self-dual point $1/\sqrt{2}$, as well as the infinity of other scalars at their origin. 

It is a long-standing problem to better understand the nature of the condensed phase, with negative potential and dilaton tadpole.  In Refs~\cite{AK,ADK}, it is proposed to be associated with an underlying non-critical string, with linear dilaton background. 
It would be interesting to compare the gauge symmetries, masses and interactions arising in supergravity and string theory to provide further evidence for such a conjecture. In addition, one may extremize the full effective action instead of the potential, in order to derive a dynamical transition between the condensed and the no-scale phases. Solving the equations of motion in Euclidean time may also yield instantonic transitions.   

A question  tackled in the core of our work is the determination of the structure constants of $\N=4$ gauged supergravity that are appropriate for describing the low energy physics of a string theory no-scale model. When $\N=4$ supersymmetry in 4 dimensions is exact,  the constants $f_{RST}$ are related to the charges of all (light) vector multiplets under the  $U(1)_{\rm L}^6\times U(1)^{22}_{\rm R}$ Cartan subgroup. In other words, they are nothing but the generalized momenta of the Narain lattice, $(p_{\rm L},p_{\rm R})\in \Lambda_{(6,22)}$~\cite{GP1,GP2}. However,  $(p_{\rm L},p_{\rm R})$ is no longer universal among the degrees of freedom of a vector multiplet, when $\N=4$ is spontaneously broken. For this reason, we have restricted the effective supergravity to a single (and potentially tachyonic) real scalar in each vector multiplet, in order to avoid any ambiguity in the choice of charge  $(p_{\rm L},p_{\rm R})$. The remaining structure constants responsible for the non-Abelian gauging of the $\N=4$ graviphotons have been determined for the tachyonic mass spectrum of the underlying string model to be reproduced. Clearly, it would be very interesting to generalize the analysis of Refs~\cite{GP1,GP2} to the case of a (total) spontaneous breaking of $\N=4$ supersymmetry, in order to identify all structure constants $f_{RST}$ from pure string theory quantum numbers.  


\section*{Acknowledgement}
 
We are grateful to Ignatios Antoniadis and Jean-Pierre Derendinger  for fruitful discussions. The work of H.P. is partially supported by a Royal Society International Cost Share Award. 


\section*{Appendix}
\label{A0}
\renewcommand{\theequation}{A.\arabic{equation}}
\renewcommand{\thesection}{A}
\setcounter{equation}{0}

For the present work to be self-contained, let us collect basic features satisfied by indefinite special orthogonal or unitary groups, and their cosets.  

\noindent $\bullet$ We are mostly interested in the group manifolds 
\be
{SO(p,q)\over SO(p)\times SO(q)},\qquad p,q\in \natural,
\label{sopq}
\ee
which are encountered in the description of closed string moduli spaces.
The $O(p,q)$ group is the set of $(p+q)\times(p+q)$ real matrices $M$ satisfying
\be
\forall X,Y\in\R^{p,q}, \quad (MX)^t\eta(MY)=X^t\eta Y,
\label{de-f}
\ee
where ${}^t$ denotes the transpose operation and $\eta=\mbox{diag}(-1,\dots,-1,1,\dots,1)$, with $p$ entries $-1$. The above definition is equivalent to saying that 
\be
\forall U,V\in \{1,\dots, p+q\}, \quad \eta_{ST}{M^{S}}_U{M^{T}}_V=\eta_{UV},
\label{defeq}
\ee
which implies the matrices $M$ to depend on 
\be
d_{p,q}=(p+q)^2-{(p+q)(p+q+1)\over 2}={(p+q)(p+q-1)\over 2}
\ee
 parameters. Because Eq.~(\ref{defeq})  yields $\det M=\pm 1$, restricting to matrices of determinant 1 imposes only a discrete condition. Hence, the dimension of $SO(p,q)$ is also $d_{p,q}$, while that of the quotient group
\be
{SO(p,q)\over  SO(q)}
\ee
is  
\be
d_{p,q}-d_{0,q}={p\over 2}\,(p+2q-1).
\ee 
A  parameterization of this manifold is given by the following subset of the equations~(\ref{defeq}), 
\be
\forall a,b\in \{1,\dots, p\}, \quad \eta_{ST}{M^{S}}_a{M^{T}}_b=-\delta_{ab}.
\label{defeq2}
\ee
To show this, we first observe that the dimension of the space of solutions of the above system is $d_{p,q}-d_{0,q}$. Next, let us view the ${M^S}_a$'s as the 
$p+q$ entries of $p$ vectors $M_a$. It turns out that the $M_a$'s can be generated by the action of $SO(p,q)$ modulo $SO(p)$. This can be seen by first defining $p$ vectors $v_a\in \R^{p,q}$ by $v_a^S=\delta^S_a$. They are invariant under the action of $SO(q)$ in the following sense: 
\be
\forall {\cal N}\in SO(q), \quad v_a={\cal H}v_a, \quad \where \quad  {\cal H}= \begin{pmatrix}I_p& 0\\0&{\cal N}\end{pmatrix}\in SO(p,q),
\ee
and where $I_p$ is the identity matrix in $p$ dimensions.
Then, there exists ${\cal M}\in SO(p,q)$ such that $M_a={\cal MH}v_a$, since  
\be
M^t_a\eta M_b=v^t_a\eta v_b=-\delta_{ab},
\ee
where the first equality follows from Eq.~(\ref{de-f}).

The dimension of the group manifold~(\ref{sopq}) is 
\be
d_{p,q}-d_{0,q}-d_{p,0}=pq.
\ee  
In the following, we specialize to the case $p=2$, which is mostly encountered in the core of the paper. Defining
\be
\hat \phi^S={1\over 2} \big({M^S}_1+i{M^S}_2\big), \quad\;\; S\in\{1,2,\dots,2+q\}, 
\label{phiZ}
\ee
the defining equations~(\ref{defeq2}) of $SO(2,q)/SO(q)$  can be written as 
\be
\begin{aligned}
&|\hat \phi^1|^2+|\hat\phi^2|^2-\sum_{i=1}^{q}|\hat\phi^{2+i}|^2 ={1\over 2},\\
&(\hat\phi^1)^2+(\hat\phi^2)^2-\sum_{i=1}^{q}(\hat\phi^{2+i})^2=0.
\end{aligned}
\ee
An explicit solution to the above constraints is given in terms of an angle $\theta$ and $q$ complex variables $\omega_i$, 
\be
\begin{pmatrix}\hat \phi^1\\\hat \phi^2\end{pmatrix} = \begin{pmatrix}\cos \theta&\sin\theta\\ -\sin\theta& \cos \theta \end{pmatrix}\!\begin{pmatrix}\phi^1\\\phi^2\end{pmatrix},\qquad \hat\phi^{2+i}=\phi^{2+i},\; \;\; i\in\{1,\dots,q\},
\ee
with the following definitions~\cite{Cala-Vesen1,Cala-Vesen2} 
\be
\begin{aligned}
\phi^1&={1\over 2\sqrt{Y}}\left(1+\sum_{j=1}^q\omega_j\right)\\
\phi^2&={i\over 2\sqrt{Y}}\left(1-\sum_{j=1}^q\omega_j\right)\\
\phi^{2+i}&={\omega_i\over \sqrt{Y}},\quad\;\; i\in\{1,\dots, q\},\\
\where \quad \;\;Y&=1-2\sum_{j=1}^q|\omega_j|^2+\Big|\sum_{j=1}^q\omega^2_j\Big|^2>0.
\end{aligned}
\label{soCV}
\ee
Hence, representatives of the classes of equivalence associated with the $SO(2)$ modding are obtained by setting $\theta=0$. The $q$ complex variable $\omega_i$ thus provide a parameterization of the quotient~(\ref{sopq}) for $p=2$.  
When $q=2$, $\phi^1,\dots,\phi^4$ can be expressed in terms of  variables $\T,\U$ instead of $\omega_1,\omega_2$, as indicated in Eq.~(\ref{TU}).

\noindent $\bullet$ Let us proceed with the coset manifold 
\be
{SU(1,1)\over U(1)}.
\label{su(11)}
\ee
The $U(1,1)$ group is the set of $2\times 2$ complex matrices $M$ satisfying 
\be
\forall X,Y\in\complex^{1,1}, \quad (MX)^\dagger\eta(MY)=X^\dagger\eta Y,
\label{su11def}
\ee
where $\eta=\mbox{diag}(-1,1)$. Because this equation amounts to having
\be
\forall U,V\in \{1,2\}, \quad \eta_{ST}{{\bar M}{}^{S}}_U{M^{T}}_V=\eta_{UV},
\label{defeq211}
\ee
the real dimension of $U(1,1)$ is 4 and the determinant of $M$ is an arbitrary phase, $\mbox{$|\det M|=1$}$. As a result, the special indefinite  unitary group $SU(1,1)$ of matrices of determinant~1 is of dimension $4-1=3$. Notice that  the ${M^S}_1$'s, $S=\{1,2\}$, can be viewed as the components of vectors $M_1\in \complex^{1,1}$ satisfying Eq.~(\ref{defeq211}) for $U=V=1$, 
\be
-|{M^1}_1|^2+|{M^2}_1|^2=-1.
\label{ty}
\ee
They define a manifold of real dimension 3, which is nothing but $SU(1,1)$ since the vectors $M_1$ can be generated by the action of $SU(1,1)$ on a constant vector $v\in \complex^{1,1}$ defined by $v^S=\delta^S_1$. This is because there is always a matrix ${\cal M}\in SU(1,1)$ such that $M_1={\cal M}v$, since 
\be
M_1^\dagger \eta M_1=v^\dagger \eta v=-1,
\ee
where we have used Eq.~(\ref{su11def}). The solutions of Eq.~(\ref{ty}) can be parameterized by unconstrained variables, which are an angle $\theta$ and a complex scalar $\S$,
\be
{M^S}_1=\varphi_S \, e^{i\theta},\quad \;\;S\in\{1,2\},
\ee
where we have defined
\be
\varphi_1= {1-i\S\over \sqrt{-2i(\S-\bar \S)}}, \quad \;\;\varphi_2= -{1+i\S\over \sqrt{-2i(\S-\bar \S)}},  \quad\;\; \Im \S>0.
\ee
As a result, representatives of the classes of equivalence associated with the $U(1)$ modding of $SU(1,1)$ are obtained by keeping fixed $\theta$, implying $\S$ to be a complex coordinates on the coset~(\ref{su(11)}). In the core of the paper, we make use of the definition~\cite{sugra1,sugra2} 
\be
\bar \Phi= \varphi_1e^{i\theta}-\varphi_2e^{i\theta}\quad \Longrightarrow\quad |\Phi|^2={2i\over \S-\bar \S}.
\ee


\bibliographystyle{unsrt}


\end{document}